\documentclass[aps,prc,onecolumn,longbibliography,showpacs,superscriptaddress,nofootinbib]{revtex4-1}
\usepackage{epsfig}
\usepackage{color}
\usepackage{hyperref}
\usepackage{mathbbol}
\usepackage{amsmath}
\usepackage{amssymb}

\newcommand{\bs}[1]{\ensuremath{\boldsymbol{#1}}}
\newcommand{\eodim}{B_{\text{\tiny D}}^{}}
\newcommand{\eotet}{B_{\alpha}^{}}
\newcommand{\eotri}{B_{\text{\tiny T}}^{}}
\newcommand{\eonn}{B_{\text{\tiny nn}}^{}}
\newcommand{\ifm}{\text{fm}^{-1}}
\newcommand{\rgm}{$\mathbb{R}$GM}

\newcommand{\mn}{m_\text{\tiny N}}
\newcommand{\mpi}{$m_\pi$}
\newcommand{\ecm}{E_\text{\tiny c.m.}}
\newcommand{\ie}{\textit{i.e.}}

\newcommand{\sgmvec}{\ensuremath{\boldsymbol{\sigma}}}

\newcommand{\anps}{^1a_{\rm np}}
\newcommand{\anpt}{^3a_{\rm np}}

\newcommand{\andd}{^2a_{\rm nD}}
\newcommand{\andq}{^4a_{\rm nD}}

\newcommand{\eftnopi}{\mbox{$\pi\text{\hspace{-5.5pt}/}$EFT}}
\newcommand{\eftpihi}{\mbox{$\pi\hspace{-11pt}\nearrow$EFT}}
\newcommand{\cpt}{$\chi$PT}

\begin{document}
\title{Spectra and Scattering of Light Lattice Nuclei from Effective Field 
Theory}
\author{J.~Kirscher}
\affiliation{Racah Institute of Physics, The Hebrew University, Jerusalem 91904, Israel}

\author{N.~Barnea}
\affiliation{Racah Institute of Physics, The Hebrew University, Jerusalem 91904, Israel}

\author{D.~Gazit}
\affiliation{Racah Institute of Physics, The Hebrew University, Jerusalem 91904, Israel}

\author{F.~Pederiva}
\affiliation{Physics Department, University of Trento, via Sommarive 14, I-38123 Trento, Italy}
\affiliation{INFN-TIFPA Trento Institute for Fundamental Physics and Applications, Trento, Italy}

\author{U.~van~Kolck}
\affiliation{Institut de Physique Nucl\'eaire, CNRS/IN2P3, 
Universit\'e Paris-Sud, F-91406 Orsay, France}
\affiliation{Department of Physics, University of Arizona,
Tucson, AZ 85721, USA}

\date{\today}
\begin{abstract}
An effective field theory is used to describe light nuclei, calculated from 
quantum chromodynamics on a lattice at unphysically large pion masses. 
The theory is calibrated at leading order 
to two available data sets on two- and three-body nuclei for two pion masses.
At those pion masses we predict 
the quartet and doublet neutron-deuteron scattering lengths,
and the alpha-particle binding energy. For \mpi=510~MeV 
we obtain, respectively,
\mbox{$\andq=2.3\pm 1.3~$fm}, \mbox{$\andd=2.2\pm 2.1~$fm}, and
\mbox{$\eotet=35\pm  22~$MeV}, 
while for \mbox{\mpi=805~MeV}
\mbox{$\andq=1.6\pm 1.3~$fm}, \mbox{$\andd=0.62\pm 1.0~$fm}, and
\mbox{$\eotet=94\pm 45~$MeV} are found. 
Phillips- and Tjon-like correlations to the triton binding energy are 
established. Higher-order effects on the respective correlation bands
are found insensitive to the pion mass.
As a benchmark, we present results for the physical pion mass,
using experimental two-body scattering lengths and the
triton binding energy as input.
Hints of subtle changes in the structure of the triton and alpha particle 
are discussed.

\end{abstract}
\maketitle

\section{Introduction}

The vast number of phenomena of the nuclear chart 
depend on a relatively small
set of 
quantum chromodynamics (QCD) parameters --- 
in the low energies relevant for nuclear physics,
a mass scale $M_{\rm QCD}$ 
associated to the strong coupling constant, the masses $m_q$
of the 
two lightest quarks, the electromagnetic coupling strength,
and the vacuum angle.
Lattice QCD (LQCD) is a numerical framework which 
enables us, at least in principle, to 
relate nuclear and QCD parameters,
once effects due to finite lattice spacing and size are removed.
The last few years have witnessed significant progress in predicting the 
properties of light nuclei with
nucleon number $A\le 4$, but at relatively large quark masses
and neglecting time-reversal and isospin violation.
(See Ref. \cite{Beane:2010em} for a 
review and a list of relevant references). 

Increasing $A$ at fixed quark masses presents significant difficulties
because the noise-to-signal rate increases exponentially. 
Although there seem to be ways around this problem \cite{Beane:2010em},
large $A$ also requires that longer
distances be covered by the lattice, since the nuclear
volume increases with $A$. As in other areas of physics,
it is profitable to change to a more effective description, in
this case to an effective field theory (EFT) involving nucleons
as degrees of freedom. Because an EFT is based on the most general
Hamiltonian with the appropriate symmetries, it is guaranteed to produce
$S$-matrix elements consistent with the $S$ matrix of the underlying theory
\cite{Weinberg:1978kz},
here QCD. 
After matching the EFT amplitudes to the LQCD-calculated
quantities at small $A$, one can describe the longer-distance dynamics involved
in larger-$A$ systems within the EFT \cite{Barnea:2013uqa},
which is considerably simpler than doing so directly within LQCD.

Most LQCD results so far concern binding energies,
but reactions convey much more information
and will command increasing attention in the years to come.
Unfortunately, as discussed in Ref. \cite{Briceno:2014tqa},
which also summarizes the progress in this field,
volume artifacts are more pronounced.
EFT quite naturally accounts for scattering
states, and allows bound states and scattering to be treated on equal footing.
Here we elaborate on the findings
of Ref. \cite{Barnea:2013uqa} for $A\le 4$ 
and extend, for the first time, LQCD predictions to 
reactions involving nuclei.
As an example, we consider neutron-deuteron (nD) scattering
at low energies, where the two $S$-wave channels
--- with total spin $s=3/2$ (quartet) and $s=1/2$ (doublet) ---
are most important.

The noise-to-signal rate in LQCD
also increases with decreasing $m_q$.
Results obtained with 
unphysical $m_q$ can, in principle,
be extrapolated to the 
physical point in a systematic way 
using chiral effective field theory ($\chi$EFT),
as long as pion masses are within the radius of convergence of the latter.
{}From $\chi$EFT with up to one nucleon ---
that is, chiral perturbation theory (\cpt) --- one obtains
the $m_q$ dependence of, for example, the average pion mass ($m_\pi$) 
\cite{Gasser:1983yg},
and of the nucleon ($\mn$) and Delta ($m_\Delta$) masses \cite{Jenkins:1991ts}.
The $m_q$ dependence of some few-nucleon observables has
also been estimated \cite{Beane:2001bc},
but unfortunately significant uncertainties still exist
due to subtleties in the proper accounting of renormalization-group (RG)
invariance in this non-perturbative context 
\cite{Beane:2001bc,Nogga:2005hy,PavonValderrama:2005uj}.



The average pion mass $m_\pi$ is commonly used as a measure for the detuned value of the average quark mass.
At present, LQCD can be carried out in the meson and single-hadron sector
down to values of $m_\pi$ close to physical,
where the low-lying mass spectrum is reproduced within theoretical error bars 
(see Ref. \cite{Thomas:2014dpa} for a status report).
Comparison with LQCD data suggests that \cpt~converges
for pion masses no larger than about 500 MeV \cite{Durr:2014oba}.
In contrast, the quark masses employed in current nuclear LQCD are likely 
beyond reach of $\chi$EFT.


As proposed in Ref. \cite{Barnea:2013uqa} and elaborated here, 
the EFT that describes existing
light-nuclear LQCD data need not include pions explicitly.
In fact, it has been understood for over fifteen years that even
at the physical pion mass
light nuclei are well described by pionless EFT (\eftnopi),
an EFT with non-relativistic nucleons interacting
through contact forces
with an increasing number of derivatives
--- each with a strength parameter or 
``low-energy constant'' (LEC) --- which contribute at increasing orders.
In two-nucleon scattering, \eftnopi~reproduces 
\cite{Kaplan:1998tg,vanKolck:1998bw,Chen:1999tn,Kong:1999sf}
the effective range expansion (ERE):
scattering lengths at leading order (LO), effective ranges
at next-to-leading order (NLO), {\it etc.}
It thus also gives two-nucleon binding momenta in
the $^3S_1$ and $^1S_0$ with corresponding accuracy.
More importantly, \eftnopi~offers a consistent extension of the ERE to other 
systems \cite{Bedaque:1997qi}.
For example, $S$-wave 
nD scattering in the 
quartet channel can be very accurately postdicted 
\cite{Bedaque:1997qi,Bedaque:1998mb,Bedaque:1999vb,Griesshammer:2004pe,Vanasse:2013sda}
once the two-nucleon LECs have been fixed in the two-nucleon system.
In the 
doublet channel, in contrast, RG invariance requires
that the three-body force with no derivatives appear already
at LO, with isospin-symmetric corrections starting beyond NLO 
\cite{Bedaque:1998kg,Bedaque:1998km,Bedaque:1999ve,Bedaque:2002yg,Griesshammer:2004pe,Griesshammer:2005ga,Ando:2010wq,Ji:2012nj,Vanasse:2014kxa}.
Current evidence from the RG in the four-body system
suggests that there is no four-body force up to NLO 
\cite{Platter:2004he,Platter:2004zs,Hammer:2006ct,Kirscher:2009aj}.
The existence of a single three-body parameter up to NLO,
which determines the three- and higher-body spectra, 
leads to many correlations among few-body observables at fixed two-body input.
Examples are the so-called Phillips \cite{Phillips:1968zze} and 
Tjon \cite{Tjon:1975} lines
obtained in plots of 
the doublet nD scattering length \cite{Bedaque:1999ve,Bedaque:2002yg} 
and
alpha-particle binding energy \cite{Platter:2004zs,Kirscher:2009aj} 
as functions of the triton binding energy.
Higher partial waves in three-nucleon scattering 
\cite{Gabbiani:1999yv,Vanasse:2013sda},
four-nucleon scattering \cite{Kirscher:2011uc},
and even $^6$Li \cite{Stetcu:2006ey} can also be 
reasonably well described in \eftnopi.

We will show 
that an analogous
approach to describe light nuclei 
is equally useful at larger $m_\pi$. 
Using higher-than-physical
$m_q$ not only increases \mpi, but also changes the nucleon mass $\mn$ 
and the masses of all other hadrons.
We will argue on the basis of scales inferred from LQCD data
that nucleons are sufficient for momenta up to $m_\pi$, with neither 
explicit pions nor other baryons.
Whether it is indeed $m_\pi$ (instead of, say, $m_\Delta-\mn$)
that determines the convergence rate of the theory used here 
is the subject of an upcoming investigation.
At each value of $m_\pi$ a pionless EFT exists with specific values of
the LECs; we refer to \eftnopi~with varying $m_\pi$ as \eftpihi~in the 
following.
Until nuclear LQCD calculations are extended to include time-reversal and 
isospin violation,
$m_\pi$ is the only QCD parameter determining nuclear properties.
Existing data 
at $m_\pi=805$ \cite{Beane:2012vq} and 510 \cite{Yamazaki:2012hi} MeV  give 
$A\le 4$ binding energies
that are much larger than in the real world and increase with the pion mass.
The dineutron is bound,
which could signal qualitative new features in lattice worlds.
An obvious question is the extent to which properties of \eftnopi~survive
in \eftpihi, where all scales change.

In Ref. \cite{Barnea:2013uqa} the binding energies of nuclei
with $A\leq 6$ were studied in LO \eftpihi~using  as input the LQCD data 
for dineutron, deuteron and triton/helion at
$m_\pi=805~$MeV \cite{Beane:2012vq}.
The alpha-particle binding energy provided a consistency check
between \eftpihi~and LQCD data, and the $A=5,6$ binding energies
obtained with \eftpihi~
can be viewed as an extrapolation of LQCD.
Here, we extend \eftpihi~
to the $m_\pi=510~$MeV LQCD data \cite{Yamazaki:2012hi}
and to a broader range of observables including scattering 
amplitudes.

The methods of \eftnopi~have for some time been deployed
in the study of reactions directly on the lattice 
\cite{Kreuzer:2008bi,Briceno:2014tqa}.
Both two-nucleon elastic scattering \cite{Beane:2013br}
and neutron radiative capture on the
proton \cite{Beane:2015yha} have been considered directly on the lattice.
Our strategy is, instead, to analyze reactions
outside the lattice box with \eftpihi~once its LECs have been determined
from binding energies at LO and,
eventually, also two-nucleon scattering observables at NLO.
We exploit the dramatic advances in the development of the so-called
{\it ab initio} methods that have taken place in nuclear physics
over the same period in which \eftnopi~was formulated.
In particular, here we employ
the effective-interaction hyperspherical-harmonic (EIHH) 
method \cite{Barnea:1999be,Barnea:2001ak,Barnea:2010zz},
and the refined resonating-group (\rgm) method \cite{Hofmann:1986}.
Although these methods have been developed for traditional
nuclear potentials, they can be adapted to pionless EFT,
as already done for \eftpihi~in the EIHH \cite{Barnea:2013uqa} 
and \eftnopi~in the \rgm~\cite{Kirscher:2009aj,Kirscher:2011uc}.

Thus, we show that \eftnopi~remains useful in nuclear systems with $A\leq4$ and 
extrapolate LQCD data to observables
that might not be as easily obtained in the lattice.
This is analogous to the use of \eftnopi~correlations 
\cite{Kirscher:2011zn,Hammer:2014rba}
to infer values of poorly measured observables in the real world.
If and when scattering observables are determined directly on the lattice, 
our predictions will be a further test of
the consistency between \eftpihi~and LQCD,
establishing
the validity of a theory with only contact interactions over a range
of $m_\pi$ from 140~MeV up to 805~MeV.
%
Such a consistency would provide a benchmark for the extension
of this method to the less-understood $\chi$EFT, once 
LQCD data reaches sufficiently small pion masses.


We summarize the article as follows. 
In Sec. \ref{sec_PEFT} we discuss the degrees of freedom 
and breakdown scale of \eftpihi~for $m_\pi$ up to $\sim 800$ MeV.
Still in Sec. \ref{sec_PEFT}, we present the LO Hamiltonian
and the regulator we use in calculations,
which employ the
computational tools introduced in Sec. \ref{sec_tool}:
the EIHH and \rgm~methods.
In Sec. \ref{sec_calib}, we determine the LO LECs
from the LQCD data for $A\le 3$
in the alternate reality assessed via LQCD at various $m_\pi$.
With the Hamiltonian thus calibrated,
we calculate in Sec. \ref{sec_res}
the 
alpha-particle binding energy, 
establish the heavy pion Phillips and Tjon lines, and predict
the doublet and quartet neutron-deuteron scattering lengths $\andd$
and $\andq$.
As we conclude in Sec. \ref{sec_conc}
the procedure is analogous to the 
development of \eftnopi~over the last decades,
namely, a calibration of a small set of parameters to data in order, 
first, to obtain
predictions of low-energy 
observables and, second, to explain correlations amongst them.



\section{Pionless Effective Field Theory}
\label{sec_PEFT}

At physical $m_\pi$, pionless EFT with nucleons as the sole degrees 
of freedom has proved useful for light nuclei in the low-momentum regime 
--- see Refs. \cite{Bedaque:2002mn,Platter:2009gz}
for reviews and Ref. \cite{Phillips:2002da} for a pedagogical introduction. 
Its organizational scheme (``power counting'') is based on
two basic scales: the breakdown scale
$Q_\text{\tiny{high}}$
estimated as $m_\pi$ and an  unnaturally small scale
$\aleph$ related to the inverse
of the two-nucleon
scattering lengths $^{1,3}a_{\rm N N}$ in the singlet/triplet channels.

For external relative momenta $k\lesssim m_\pi/2$, 
the deuteron and the virtual singlet bound state are
the only singularities of the two-body scattering amplitude. 
All mesons and excited baryons are short-range effects.
The amplitude from a Lagrangian built of 
derivative contact operators made
of nucleon fields can be matched to all orders of 
the ERE.
Matching the LO amplitude to the ERE
results in a
$C_{s,t}\propto 4\pi (^{1,3}a_{\rm NN})/\mn \sim 4\pi/(\mn\aleph)$ scaling
of the non-derivative four-nucleon contact term.
As the scattering lengths $^{1,3}a_{\rm NN}$ are
large relative to the pion range $1/$\mpi~for both
NN $S$-wave channels, a refined power counting is required
\cite{Kaplan:1998tg,vanKolck:1998bw}
that goes beyond naive dimensional analysis.
Of course,
care has to be taken
that the necessary regularization and the 
inclusion of higher-order contributions
do not introduce poles within the radius of convergence.
As long as those singularities 
are beyond the pion threshold, \mpi, \eftnopi~converges for two-nucleon 
processes at momenta $Q< Q_\text{\tiny{high}}$,
including the $^3S_1$ (deuteron) 
and $^1S_0$ poles \cite{Chen:1999tn,Kong:1999sf}.
In LO the two LECs $C_{s,t}$ suffice.

Extending \eftnopi~to systems with more nucleons requires understanding
how $\aleph$ enters the LECs of multinucleon interactions.
The fact that the non-derivative six-nucleon contact interaction
is needed to define the EFT at LO \cite{Bedaque:1998kg,Bedaque:1998km}
implies its LEC scales as $D_{d}\sim (4\pi)^2/(\mn\aleph^4)$.
In contrast, the apparent lack of similar RG enhancements in other
contact interactions suggests they appear only in higher orders.

As \eftnopi~is applied beyond the deuteron, one needs to account
for effects of the Coulomb force among protons.
The importance of Coulomb effects is characterized by a ratio $\alpha m_N/Q$,
where $\alpha$ is the fine-structure constant.
Although crucial for very low-energy proton-proton \cite{Kong:1999sf} and 
proton-nucleus scattering, the Coulomb interaction should be subleading 
in relatively deep ground states such as helion and alpha particle, where
binding momenta are much larger than $\alpha m_N$.
\par
At LO, the \eftnopi~Lagrangian can be written as
\begin{widetext}
\begin{eqnarray}\label{eq_lo-lagrangean}
\mathcal{L}_{LO}&=&N^\dagger\left(i\partial_0+\frac{\vec{\nabla}^2}{2\mn}\right)N 
+\frac{C_{s}}{8}\left(N^T\sigma^2\sigma^i\tau^2\,N\right)^\dagger
\left(N^T\sigma^2\sigma^i\tau^2\,N\right) 
+ \frac{C_{t}}{8}\left(N^T\sigma^2\tau^2\tau^a\,N\right)^\dagger
\left(N^T\sigma^2\tau^2\tau^a\,N\right)
\nonumber\\
& &  + D_{d}  (N^\dagger N)(N^\dagger N)(N^\dagger N),
\end{eqnarray}
\end{widetext}
where $N$ is a bi-spinor in both spin and isospin spaces, and
$\sigma^i$ ($\tau^a$) are the spin (isospin) Pauli matrices,
the index $i$ ($a$) running over spin (isospin) vector components of the 
projection operators on the spin singlet (triplet) state.
Higher orders contain terms with more derivatives and/or nucleon fields,
including those necessary to ensure Lorentz invariance (in a $Q/\mn$ expansion).

Somewhat surprisingly, \eftnopi~seems to converge 
for triton and helion \cite{Bedaque:1999ve,Bedaque:2002yg,Ando:2010wq},
and even for the more-bound alpha particle
\cite{Platter:2004zs,Kirscher:2009aj}.
At the physical point, 
\eftnopi~is useful even at 
LO to explain features like correlations amongst three-body observables (the Phillips line)
and between three- and four-body data (the Tjon line),
with just the 
neutron-proton scattering lengths 
$^{1,3}a_{\rm np}$
as input.
With an additional condition which conventionally fixes either 
the triton binding energy $\eotri$ or the
neutron-deuteron doublet scattering length $\andd$, 
a few four-nucleon observables
--- {\it e.g.}, the binding energy of $^4$He
\cite{Platter:2004zs,Kirscher:2009aj},
and the neutron-triton and 
proton-helion scattering lengths~\cite{Kirscher:2011uc} ---
have been found to agree with data
within the expected uncertainty margin.
The only exception so far seems to be
the resonance location in the $0^-$ neutron-triton channel,
which was found to be 
cutoff, and thus
renormalization-scheme dependent~\cite{Deltuva:2011}.
The origin of this pathology is unknown.
LO results for $^6$Li \cite{Stetcu:2006ey} do not allow conclusions
about the range in $A$ where \eftnopi~converges.


With the usefulness at physical \mpi~thus established,
we follow an analogous approach at heavier $m_q$.
Available lattice data~\cite{Thomas:2014dpa} identifies the pion, still, 
as the lightest meson and the 
Delta as the lowest excited state of the nucleon.
However, the ratios between the nucleon, pion, and 
Delta masses change, see Table~\ref{tab.scales}.
Also, nuclei become increasingly more bound.

\begin{table*}
\renewcommand{\arraystretch}{1.1}
\caption{\label{tab.scales}{\small Relevant scales for a low-energy 
nuclear effective field theory. Physical data in the first column 
is relevant for \eftnopi,
lattice data summarized in the second and third for \eftpihi.
All numbers are given in MeV.}}
\begin{tabular}{ccccc}
\hline\hline
\hline
pion mass          & \mpi   & $139.5\pm 0.1$~\cite{Agashe:2014kda}  
& $511\pm 2$ \cite{Yamazaki:2012hi}    & $806\pm 1$ \cite{Beane:2012vq} \\ 
\hline
nucleon mass       & $\mn$  & $939\pm 1,938\pm 1$~\cite{Mohr:2012tt} 
& $1320\pm 3$  \cite{Yamazaki:2012hi}  & $1634\pm 18$ \cite{Beane:2012vq} \\ 
Delta-nucleon mass difference & $\delta_\Delta=m_\Delta-\mn$ 
& $292\pm 1$~\cite{Anisovich:2005tf} 
& $\approx 200$~\cite{Leinweber:1999ig} 
& $\approx 180$~\cite{Leinweber:1999ig} \\
dineutron binding energy  &$\eonn$   & ---  & $7.4\pm 2$ \cite{Yamazaki:2012hi}   
& $15.9\pm 4$  \cite{Beane:2012vq} \\
deuteron  binding energy  &$\eodim$  & $2.22$~\cite{Greene:1986vb}  
& $11.5\pm 2$ \cite{Yamazaki:2012hi}  & $19.5\pm 5$  \cite{Beane:2012vq} \\
triton  binding energy    &$\eotri$  & $8.482$~\cite{Wapstra:1985zz} 
& $20.3\pm 4.5$ \cite{Yamazaki:2012hi} & $53.9\pm 10.7$ \cite{Beane:2012vq} \\ 
\hline
inverse singlet scattering length& $\anps^{-1}$     & $-8.31$~\cite{Dumbrajs:1983jd}    
& n.a. & $84.7\pm 18$  \cite{Beane:2013br}  \\
inverse triplet scattering length& $\anpt^{-1}$     & $36.4$~\cite{Dumbrajs:1983jd}     
& n.a.  & $108\pm 13$   \cite{Beane:2013br} \\
\hline
Delta effective momentum & $\sqrt{2\mn\delta_\Delta}$ & $741$     & $890$        
& $767$ \\
two-nucleon binding momentum &$\sqrt{\mn (\eonn+\eodim)/2}$ 
& $46$  & $112$ & $170$  \\
triton-to-deuteron binding ratio & $\eotri/\eodim$ & $3.82$ & $1.8$ & $2.8$\\
\hline\hline
\end{tabular}
\end{table*}

The relevant momentum is very clear in the two-nucleon system,
from either the inverse scattering lengths or the two-nucleon
binding momentum estimated from the average two-nucleon binding energy.
At all pion masses it is much smaller than the nucleon mass,
meaning nucleons are nonrelativistic,
and even than the pion mass itself, ensuring pions can be integrated out.
However,
in contrast to the physical world, $m_\pi>m_\Delta-\mn\equiv \delta_\Delta$ 
for the two lattice 
simulations,
and hence one might wonder if the Delta should not be included
as an explicit degree of freedom.


The reason the Delta can still be integrated out is, of course,
that in a nonrelativistic theory the relevant quantity for convergence
is momentum,
not heavy-particle mass. 
In this case, it is the ``Delta effective momentum'' $\sqrt{2\mn\delta_\Delta}$,
which remains above, or at least near, the pion mass.
That $\sqrt{2\mn\delta_\Delta}$ is the relevant scale 
was shown explicitly in Ref. \cite{Savage:1996tb} for
the two-nucleon $^1S_0$ channel.
In this case, the lowest accessible
state with excitations has two Deltas, and in addition
to $C_s$ two other non-derivative contact interactions need to be included:
two-nucleon/two-Delta and four-Delta.
Under the assumption that all three LECs
are of a similar size $C_0$,
they scale as \cite{Savage:1996tb}
\begin{equation}\label{eq.dd_scaling}
C_0 = \frac{4\pi}{\mn}\left(^{1}a_{\rm NN}^{-1}+ \sqrt{2\mn\delta_\Delta}\right)^{-1}
      \approx (234~\text{MeV})^{-2}
\end{equation}
at \mpi=140~MeV. 
Because the inverse value of the singlet scattering length for \mpi=805~MeV,
displayed in Table~\ref{tab.scales}, is about 10 times larger
in magnitude than its physical analog, 
the ensuing size of
$C_0$ would decrease 
and allow for higher typical momenta in the 
two-nucleon amplitude. 
However, the Delta effective momentum
is still several times
larger than the inverse scattering length. 
Operators in a 
Deltaful, pionless theory 
should then show similar scaling behavior 
as for physical pion mass, where the Delta can be integrated out.
Removing the Delta generates an effective range not accounted for
in LO, but this contribution is characterized by the Delta effective momentum,
which does not seem to be smaller than the inverse pion mass.

This argument can be generalized to other channels
\cite{Cohen:2004kf}
where $\sqrt{2\mn\delta_\Delta}$ is replaced by $\sqrt{\mn\Delta_{\rm r}}$,
with $\Delta_{\rm r}$ the difference between the mass of the state 
containing the nucleon excitation(s) and $2\mn$. 
Since lattice results suggest that the lowest state with a single excitation
involves the Roper resonance, whose mass is somewhat larger than
the Delta, one does not expect a significant decrease in convergence rate
by keeping only the nucleon explicit in the EFT.


Therefore, we formulate \eftpihi~as an EFT formally equivalent
to \eftnopi, but with different scales and values for the LECs.
The breakdown or high-momentum scale
$Q_\text{\tiny{high}}$ is assumed to be the smallest of
\mpi~and $\sqrt{\mn\Delta_{\rm r}}$.
The low-momentum scale $Q_\text{\tiny{low}}$ is set by the binding momenta
of the nuclei we consider and by the external momenta
in the reactions we are interested in. 
We expand all observable in powers of 
$Q_\text{\tiny{low}}/Q_\text{\tiny{high}}$.
Eventually, an NLO calculation will yield an estimate on the convergence
rate and thereby the breakdown scale of \eftpihi.
The Lagrangian in LO is given by Eq. \eqref{eq_lo-lagrangean},
in which 
four $m_\pi$-dependent
parameters enter: the nucleon mass $\mn$ and the LECs $C_{s,t}$ and $D_d$.

For the calculation of few-body observables we solve the Schr\"odinger 
equation in configuration space. 
The potential is the sum of all irreducible contributions to the 
$A$-body scattering matrix from the Lagrangian.
This amounts at LO to the sum of three tree-level diagrams 
with vertex factors $C_{s,t}$ and $D_d$.
The infinities resulting from the zero-range contact interactions are here
regularized via Gaussian regulator functions,
$\Lambda^3 \exp(-\Lambda^2\mathbf{r}_{ij}^2/4)/(16\pi^{3/2})$ for two nucleons 
$i,j$ 
and
$\Lambda^6\exp[-\Lambda^2(\mathbf{r}_{ik}^2+\mathbf{r}_{jk}^2)/4]/(64\pi^{3})$ 
for three nucleons 
$i,j,k$,
where $\Lambda$ {\it arbitrarily} separates states included
explicitly as propagating degrees of freedom from states accounted for
implicitly in the LECs. If it is smaller than the breakdown scale it produces larger errors than the truncation of the EFT Lagrangian (Sec.~\ref{doublet_section} exemplifies ramification of a 
violation of this condition).
The resulting Schr\"odinger equation for
the $A$-body wave function $\Psi$ and the corresponding energy $E$ 
takes the form
\begin{widetext}
\begin{equation}\label{hpsi}
\left\{-\sum_i \frac{\nabla^2_i}{2 \mn} 
+ \sum_{i<j} \frac{1}{2}
\left[C_{1,0}+C_{0,1} + (C_{1,0}-C_{0,1})\,\sgmvec_i\cdot\sgmvec_j \right] 
e^{-\frac{\Lambda^2}{4}r_{ij}^2} 
+ \sum_{i<j<k}\sum_{cyc} D_1 e^{-\frac{\Lambda^2}{4}(r_{ik}^2+r_{jk}^2)}\right\} \Psi
=E\,\Psi.
\end{equation}
\end{widetext}
Here, a factor from 
the regulator was absorbed
into the bare couplings of Eq.~\eqref{eq_lo-lagrangean}: 
\begin{eqnarray}
C_{0,1}(\Lambda)&=&\frac{\Lambda^3}{16\pi^{3/2}}C_{s}(\Lambda), 
\label{C01}\\
C_{1,0}(\Lambda)&=&\frac{\Lambda^3}{16\pi^{3/2}}C_{t}(\Lambda), 
\label{C10}\\
D_1(\Lambda)&=&\left(\frac{\Lambda^3}{8\pi^{3/2}}\right)^2D_d(\Lambda).
\label{D1}
\end{eqnarray}
As in any EFT, the bare LECs depend on $\Lambda$ so as to guarantee that
observables do not.
The $\Lambda$-dependent LECs are determined from input data in Sec. 
\ref{sec_calib},
after we discuss the solution of Eq. \eqref{hpsi} in the next section.

\section{Toolbox}
\label{sec_tool}
To solve the Schr\"odinger equation we have utilized two computational methods:
EIHH and \rgm. Hereafter, we present a short description of both methods.

\subsection{The Effective-Interaction Hyperspherical Harmonics 
Method}

The hyperspherical coordinates are the $D$-dimensional 
generalization of the $3$-dimensional spherical or polar coordinates. 
As such they allow
the description of the $A$-body wave function in terms of a single length 
variable,
the {\it hyper}-radius $\rho$, and $(D-1)$ {\it hyper}-angular variables 
$\Omega$~\cite{Oehm:1990qt,Viviani:1994pm}.
Removing the center-of-mass coordinate, the $A$-body dynamics can be described 
by $A-1$ Jacobi vectors 
$\bs{\eta}_1$, ..., $\bs{\eta}_{A-1}$, therefore $D=3A-3$. 

The nice feature of these coordinates is that, in perfect analogy to the 
two-particle case, 
the kinetic energy operator $T$
of the $A$-particle system splits into
a {\it hyper}-radial and {\it hyper}-centrifugal terms, with
a {\it hyperspherical} angular momentum operator $\hat{K}$ that depends 
on $\Omega$. 
The resulting $A$-particle Hamiltonian reads
\begin{equation}\label{H^[A]}
H^{[A]} = - \frac{1}{2\mn}\left(\Delta_{\rho}-\frac{\hat{K}^2}{\rho^2}\right)
+ V^{[A]}(\rho,\Omega)\,,
\end{equation}
where $\Delta_{\rho}$ is the {\it hyper}-radial Laplacian.

The hyperspherical harmonics (HH) ${\cal Y}_{[ K]}$ are the $A$-body 
generalization of the spherical harmonics. 
As such they are the eigenfunctions of $\hat{K}^2$ with eigenvalues
$K(K + 3A-5)$. They form a complete set of {\it hyper}-angular basis functions. 
Choosing a complementary set of {\it hyper}-radial basis states $R_n(\rho)$,
the $A$-body wave function can be expanded in the form
\begin{equation}\label{HH}
\Psi(\rho,\Omega) = \sum_{n[K]} C_{n[K]}R_n(\rho) {\cal Y}_{[K]}(\Omega)
\end{equation}
with coefficients $C_{n[K]}$.
The nuclear wave function $\Psi$ must be complemented by the 
spin-isospin parts, and the whole function must be antisymmetric. 
The construction of antisymmetric HH spin-isospin basis states 
is a non-trivial task, 
which, however, has been solved in Refs. \cite{Barnea:1998zz,Barnea:1998zza}.

To accelerate the convergence rate of the HH expansion, Eq.~(\ref{HH}), 
we construct an effective interaction (EI) using the Lee-Suzuki similarity 
transformation~\cite{Suzuki:1980yp}. Applying
the this method to the HH basis  
we identify the model space $P$ with all the $A$-body HH states such 
that $K\leq K_{max}$, and
the complementary space $Q=1-P$ 
as the rest of the Hilbert space. 
The Lee-Suzuki method then gives a recipe to construct a similarity 
transformation 
such that the spectrum of the resulting effective $P$-space Hamiltonian,
$H^{[A]eff}=T+V^{[A]eff}$, coincides with 
the spectrum of $H^{[A]}$.
Finding $V^{[A]eff}$, however, is as difficult as solving the original problem, 
and
therefore we do not search for the {\it total} EI, but for a {\it partial} EI 
constructed through the solution of the simpler two- and three-body problems.

The resulting EI is tailored to our HH model space, and constrained to 
coincide with the bare one when enlarging the model space. 
This EIHH method \cite{Barnea:1999be,Barnea:2001ak,Barnea:2010zz} 
has been successfully applied to the study of bound states and reactions for
nuclear systems with $3 \leq A \leq 7$.

\subsection{The Refined Resonating-Group Method}

In contrast to the EIHH method where the few-body wave function is expanded over
a complete set of states, the \rgm~is a variational approach that utilizes 
an over-complete set of states
(for its original formulation, see Ref. \cite{Wheeler:1937zz,Wheeler:1937zza}; 
for the refinement and implementation, Ref. \cite{Hofmann:1986}).
To construct these states, the \rgm~considers all possible 
channels $\{[c]\}$, where each channel consists of a specific spin-isospin
configuration $\Xi_{[c]}$, a set of Jacobi vectors
$\bs{\eta}_1,\ldots,\bs{\eta}_{A-1}$, and the angular momentum quantum numbers
$\ell_1,\ldots,\ell_{A-1}$ associated with these vectors.
The orbital functions are given by the ansatz
\begin{equation}\label{GG_ch}
   R_{n[c]}(\bs{\eta}_1,\ldots,\bs{\eta}_{A-1})=
     \prod_{j=1}^{A-1} \eta_j^{\ell_{j}}Y_{\ell_j m_j}(\hat{\eta}_j)
                    e^{-\kappa_{n j}\eta_j^2}\,,
\end{equation}
where $Y_{\ell m}$ are the spherical harmonics, and $\kappa_{n j}$ are a set of 
width parameters used to expand the wave function, \ie, the sum over 
the channels includes an expansion of each radial
dependence in Gaussians with widths $\lbrace \kappa_{n j}\rbrace$.

The few-body wave function is then a linear combination of an antisymmetric 
product of a spin-isospin channel state and the orbital function, coupled to 
yield the desired total angular momentum quantum numbers $JM$,
\begin{equation}\label{GG}
\Psi_{JM} = {\cal A}\sum_{n[c]} C_{n[c]}\left[R_{n[c]}{\otimes}\Xi_{[c]}\right]^{JM} 
\,.
\end{equation}
The sum over channels allows the consideration of all possible spin-isospin 
configurations or clusters $\Xi_{[c]}$. In practice, however, our implementation omits channels 
that have negligible contribution to the wave function. For example, the
ansatz for the 
alpha-particle wave function includes triton-proton, 
helion-neutron, and deuteron-deuteron spin-isospin configurations. 
The conceivable two-neutron--two-proton arrangement was found to contribute 
less than $100~$keV to $\eotet$ and therefore is 
not included in the variational 
ansatz.

Thus, the \rgm~ method includes three intertwined expansions: 
{\it i)} the cluster
or resonating-group expansion, which defines the spin-isospin configuration 
and the Jacobi coordinates;
{\it ii)} the partial-wave expansion; and 
{\it iii)} the expansion in Gaussian functions. 
Convergence is assessed along each of those ``axes''. 
First, the thresholds of a system 
serve as guidance for the initial choice of resonating groups. 
Second, contributions from subleading partial waves are considered. 
For $s$-wave nuclei, and central forces, $\ell > 0$ configurations do
not have to be included due to the cluster expansion.
Consequently, at this order of our EFT we consider only $\ell=0$ terms in 
our description of the light, $A\leq 4$ nuclei.
Third, the set of Gaussians is extended and scaled until this modification
of the model space does not affect binding energies by more than $1\%$.
 
With the \rgm, we also calculate scattering observables. 
To solve the few-body problem with the \rgm~
for a range of cutoff ($\Lambda$ as introduced above to obtain the regularized Eq.~\ref{hpsi}) values, \ie, 
to approximate a wave function with structure around
$\eta_j\approx\Lambda^{-1}$, 
the variational basis has to be either very large --- leading to numerical
instabilities--- or tailored to each $\Lambda$ --- 
requiring a convergence check with regards to all 
parameters of the basis set. 

Our variational approach is analog to Kohn-Hulth\`{e}n's 
method~\cite{Kohn:1948} which minimizes a
functional parameterizing the reactance matrix, corresponding to 
Ricatti-Bessel asymptotic solutions 
for uncharged particles and Coulomb functions for charged fragments. 
We use in- and out-going waves as boundary conditions (spherical
Hankel functions $h^\pm$), because this method turned out to be more accurate 
in practice. 
Specifically, for two-fragment scattering with an incoming channel
$c$ we denote the relative intercluster Jacobi coordinate by $\eta_c$ 
and make the ansatz
\begin{equation}\label{eq.wfktans}
\Psi={\cal A}\left(-h_{c}^{-}(\eta_c)
             +\sum_{c'}S_{c c'}h_{c'}^{+}(\eta_{c'})
            +\sum_{n[c]\in\mathcal{C}}D_{n[c]}R_{n[c]}\right)\,,
\end{equation}
with variational parameters $S_{c c'}$ (the $S$ matrix) and $D_{n[c]}$.
If either target or projectile are compound objects, {\it e.g.}, 
the deuteron in Sec.~\ref{sec_3},
their wave functions are predetermined via the ansatz in Eq.~(\ref{GG}) 
and multiplied with the asymptotic 
solutions $h^\pm$ of the relative motion.
For small distances $\eta_c$, the interaction between nucleons of different 
fragments  
is non-zero and the full scattering wave function will differ from the 
asymptotic form as
given by the first two 
terms in Eq.~(\ref{eq.wfktans}). 
This difference is described by the third term in Eq.~(\ref{eq.wfktans}).
Convergence and stability are assessed with respect to the subset 
$\mathcal{C}$ which is 
taken from the full set of channels.
It is sufficient to include those $n[c]$ in $\mathcal{C}$ which are non-zero 
for $\eta_c\approx\Lambda^{-1}$ and, as Gaussians, square-integrable.
For $\eta_c\gg\Lambda^{-1}$, this expansion should be zero, \ie, $\Psi$ is 
identical to the asymptotic solution.
The Kohn-Hulth\`{e}n variational condition expressed in terms of the 
scattering matrix is
\begin{equation}\label{eq.kohnvari}
\delta\left\lbrace\langle \Psi \vert\left(H-\ecm)\right\vert \Psi \rangle
    -iS_{c c}\right\rbrace=0 \,,
\end{equation}
where  $\ecm$ is the center-of-momentum energy.
This 
condition yields optimal values for 
$S_{c c'}$ 
and $D_{n[c]}$.
Here the channel index $c$ discriminates between different two-body 
fragmentations, {\it e.g.}, neutron/deuteron or
neutron/neutron-proton singlet, and angular momentum. 
Using an appropriate decomposition of the Hamiltonian 
(for the latest summary and references to the original
work see Ref. \cite{Kirscher:2015ana}),
the variational coefficients $S_{c c'}$, $D_{n [c]}$ can be expressed in terms of 
integrals
of the short-ranged part of the potential. 
Therefore, an accurate expansion of the asymptotic solution is
required for a finite range. 
In practice, we minimize
\begin{equation}\label{eq.asyexp}
I(\epsilon)=\int_0^\infty d\eta\left(h_c^\pm(\eta)-\sum_{n[c]}C_{n[c]}R_{n[c]}\right)^2
   \,\eta e^{-\epsilon \eta^2}\,,
\end{equation}
to approximate the Hankel or Coulomb functions.
Finally, we obtain scattering lengths 
from the phase shift $\delta(\ecm)$ at a finite 
$\ecm$ through
\begin{equation}\label{eq_scatt_len}
a(\ecm)=-\frac{1}{k\cot\delta(\ecm)}\,.
\end{equation}
As the scattering length is defined for $\ecm=0$,
the uncertainty due to this approximation has to be assessed.
In this work, we extracted $a$ at $0.001$~MeV, 
used $10$-$13$ Gaussians to expand the deuteron and
singlet neutron-proton fragment in the
three-body scattering calculations, and 
adapted the Hankel functions with a weight
$\epsilon=0.03~\textrm{fm}^{-2}$.

To conclude, we summarize the convergence check:

\begin{itemize}
\item
First, we determine appropriate Gaussian basis for the fragments by fixing the 
number of Gaussians and optimize their widths via a genetic 
algorithm~\cite{Winkler:1994jg}.

\item
Second, we diagonalize the Hamiltonian, Eq. \ref{hpsi},
in the scattering basis. 
This basis uses a different coupling
scheme which adopts the one implied in Eq.~(\ref{GG}) for each fragment. 
The total fragment spins
are coupled to a channel spin which forms, with the orbital angular momentum 
on the relative coordinate
$\eta_c$ between the fragments, the total $J$. 
We enlarge $\mathcal{C}$ until the lowest eigenvalues
reproduce the thresholds defined by the ground states of the fragments and 
the bound states of the
compound system of the two fragments, 
if there is a bound state in the channel (the
triton in Sec.~\ref{sec_3}).

\item
Third, we take 
\begin{equation}\label{eq.convchecks}
\lim_{\epsilon\to 0}~I(\epsilon)\;\;\text{and}\;\;\lim_{E\to 0}~a(E)
\end{equation}
in Eqs.~(\ref{eq.asyexp} and (\ref{eq_scatt_len}).
While taking both limits, we identify a plateau in the predicted scattering 
length for
$\epsilon<\Lambda^2$ and $\ecm<0.0001~$MeV. 
\end{itemize}
After these steps, we deem the basis
large enough for an accuracy that is then dominated by the higher-order 
contributions of the EFT expansion.
\subsection{Comparison}

\begin{widetext}
\begin{figure*}[tb]\begin{center}
\includegraphics[width=1.0 \textwidth]{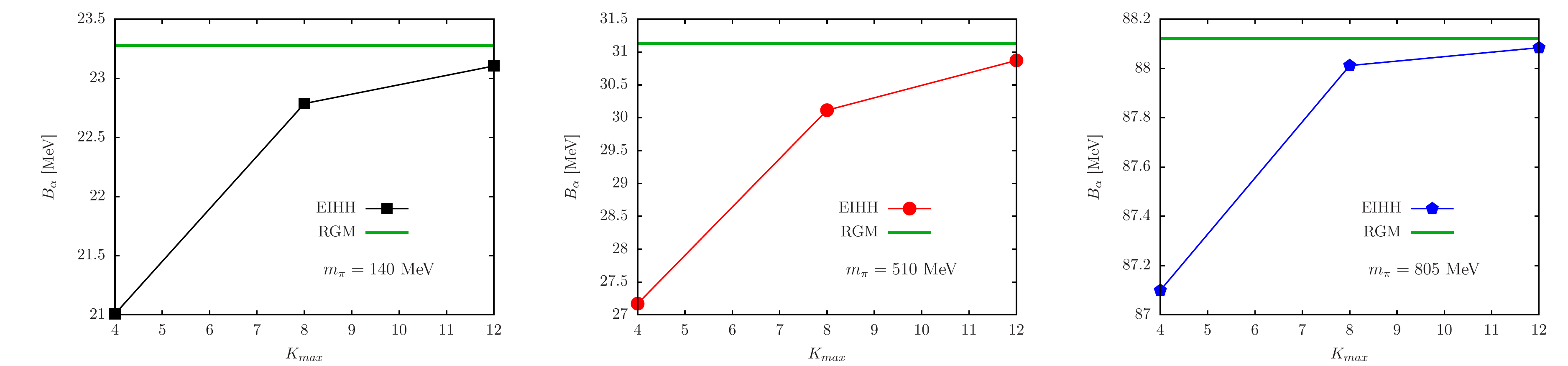}
\caption{\label{fig_rgm_eihh} Dependence of 
the alpha-particle binding energy $\eotet$ (in MeV)
calculated with the EIHH 
method on the maximal {\it hyper}-angular eigenvalue $K_{max}$. 
Results are shown for a \eftpihi~interaction with 
$\Lambda=2~\text{fm}^{-1}$
at 
\mpi = 140 MeV (left panel), 
510~MeV (center), and 
805~MeV (right). 
The horizontal green line represents the corresponding \rgm~result.}
\end{center}
\end{figure*}
\end{widetext}

With EFT parameters calibrated as described below,
we compared the results for $\eotri$ and $\eotet$ 
of the \rgm~with the corresponding EIHH values
to test the accuracy of the resonating-group expansion. 
As an example, we show in Fig.~\ref{fig_rgm_eihh}
the convergence of EIHH calculations
to the \rgm~results for $\eotet$ at a cutoff $\Lambda=2~\text{fm}^{-1}$. 
For all three pion masses, the EIHH converges with $K_{max}$ to the
respective \rgm~value.

For subsequent calculations the \rgm~was
chosen to minimize computing time. 

\section{Calibration}
\label{sec_calib}

Through Eq.~\eqref{hpsi},
all LO predictions depend on three LECs
$C_{S,T}$ ($S,T=0,1$ or $1,0$) and $D_1$,
besides the nucleon mass. 
Lattice data are available for the two- and three-body binding energies 
($\eonn,\eodim,\eotri$) at \mbox{\mpi$=510~$MeV~\cite{Yamazaki:2012hi}} 
and at \mpi$=805~$MeV~\cite{Beane:2012vq}.
At \mpi$=805~$MeV~\cite{Beane:2013br},
the singlet and triplet scattering lengths ($\anps,\anpt$)
and effective ranges are also available .
We fit $C_{S,T}$ to the 
two-nucleon binding energies in the 
singlet and triplet channels 
($\eonn,\eodim$)
by solving the Schr\"odinger equation via the Numerov algorithm.
The $D_1$ term is fixed through $\eotri$ using the \rgm.
For comparison, we also consider the physical pion mass, \mpi=140~MeV,
where we fit the experimental singlet scattering length,
in addition to experimental deuteron and triton binding energies.

\begin{widetext}
\begin{figure*}[tb]\begin{center}
\includegraphics[width=1.0 \textwidth]{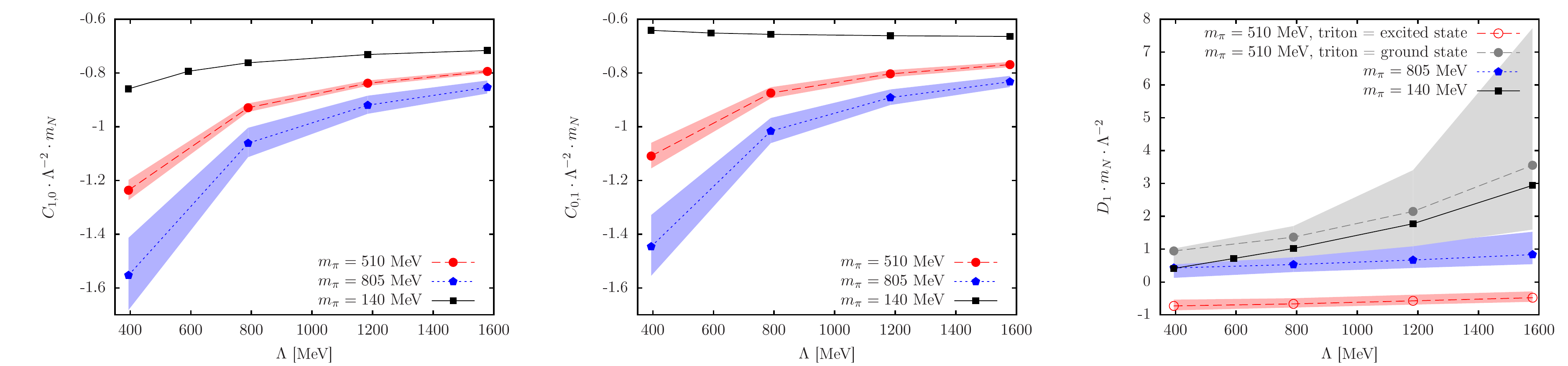}
\caption{\label{fig_lec}
Dependence on the cutoff $\Lambda$ (in MeV) of the LO LECs of \eftpihi:
$m_N \Lambda^{-2} C_{1,0}$ (left panel), $m_N \Lambda^{-2}C_{0,1}$ (middle), 
and $m_N \Lambda^{-2} D_1$ (right).
The squares (\mpi=140~MeV), circles (510~MeV), and pentagons (805~MeV) 
represent 
values fitted to the central
values of the shallowest two-nucleon $S$-matrix poles in the singlet and 
triplet channels. $D_1$ is adjusted to
the triton as the ground (full circles) or (for \mpi=510~MeV only) 
first-excited (empty circles) three-nucleon state. 
The shaded uncertainty is obtained by varying the input data within its margin 
of error.}
\end{center}
\end{figure*}
\end{widetext}

These renormalization conditions 
determine the $\Lambda$ dependence of the LECs.
The values of the bare LECs for cutoffs 
\mbox{$\Lambda=2,\;4,\;6,\;8\;\rm{fm}^{-1}$} are given in Table~\ref{tbl:LECs}.
and
a graphical representation of the fit results is given in Fig.~\ref{fig_lec}.
The input to calibrate the values at the physical pion mass 
(black squares in Fig.~\ref{fig_lec}), namely
the deuteron binding energy and the singlet neutron-proton scattering length, 
are known accurately.
Thus we abstain from a display of the sensitivity of those values to the 
uncertainty in the input.
For the unphysical pion masses, the uncertainty in the input data is 
significant.
For each cutoff, we thus obtain LECs not only for the central values but 
also for the boundaries of the two- and three-body binding energies.
In the case of $D_1$, we fix the two-nucleon LECs to their central 
values when varying $\eotri$ within its error margins.
The widths of the blue (\mpi=805~MeV) and red/gray (510~MeV) bands 
in Fig.~\ref{fig_lec}
represent
how input-data uncertainty translates into coupling strength uncertainty.

\begin{table}[bt]
\begin{center}
\caption{The 
LO LECs $C_{S,T}$ and $D_1$ [GeV] for real ($m_{\pi}=140$ MeV) and lattice
($m_{\pi}=510,\;805\;\rm{MeV}$) nuclei 
for various values 
of the momentum cutoff 
\mbox{$\Lambda$ [fm$^{-1}$]}. $D_1^{(*)}$ yields the triton
as the ground (excited) state.}
\label{tbl:LECs}
\begin{tabular}
{c@{\hspace{5mm}} c@{\hspace{5mm}} c@{\hspace{5mm}} c@{\hspace{5mm}} c@{\hspace{5mm}} c}\hline\hline
$m_{\pi}$ & $\Lambda$ &  $C_{1,0}$  &  $C_{0,1}$ & $D_1$ & $D^{*}_1$  \\
\hline
  140   & 2  &      $-$0.142  &  $-$0.106  &  0.068 & - \\
        & 4  &      $-$0.505  &  $-$0.435  &  0.677 & - \\
        & 6  &      $-$1.09  &  $-$0.986  &  2.65 & - \\
        & 8  &      $-$1.90  &  $-$1.76  &  7.82 & - \\
\hline
  510   & 2  &      $-$0.145  &  $-$0.130  &  0.157 & $-$0.120 \\
        & 4  &      $-$0.438  &  $-$0.412  &  0.907 & $-$0.441 \\
        & 6  &      $-$0.889  &  $-$0.853  &  3.21 & $-$0.855 \\
        & 8  &      $-$1.50  &  $-$1.45  &  9.44 & $-$1.27 \\
\hline
  805   & 2  &      $-$0.148 &  $-$0.138  &  0.071 & - \\
        & 4  &      $-$0.405 &  $-$0.388  &  0.354 & - \\
        & 6  &      $-$0.789 &  $-$0.766  &  1.00 & - \\
        & 8  &      $-$1.30  &  $-$1.27  &  2.22 & - \\
\hline\hline
\end{tabular}
\end{center}
\end{table}

Some aspects of the cutoff dependence of the LECs shown in Fig.~\ref{fig_lec}
can be understood from general arguments.
For a non-derivative four-nucleon LEC $C$ (multiplied by 
$\Lambda^3/(16\pi^{3/2})$
as in Eqs. \eqref{C01} and \eqref{C10}) which determines a scattering 
length $a$, 
an expansion 
in powers of relative momentum over $\Lambda$ 
of the loop integrals appearing in the $T$ matrix gives 
\cite{vanKolck:1998bw}
\begin{equation}\label{eq_c0_coord}
\mn \Lambda^{-2} C(\Lambda)=\theta_0
+\frac{\theta_1}{a\Lambda}+\mathcal{O}\left((a\Lambda)^{-2}\right)\,,
\end{equation}
where $\theta_i$ are regulator-dependent numbers of ${\cal O}(1)$
that depend neither on $a$ nor on $\mn$, and thus also not on $m_\pi$.
This large-$\Lambda$ behavior is apparent in the left and middle panels 
of Fig.~\ref{fig_lec},
where we display $\mn\Lambda^{-2}C_{S,T}$ rather than $C_{S,T}$. 
As we can see, all curves approach a limit 
$\theta_0 \simeq 0.7$,
at a rate that depends on $a$.
The different sign of the scattering length in the singlet channel
for physical \mpi~results 
in a different approach to the asymptotic value
compared to the other channels and pion masses,
where relatively shallow bound states exist.

\begin{figure}[tb]
\begin{center}
\includegraphics[width=0.5 \textwidth]{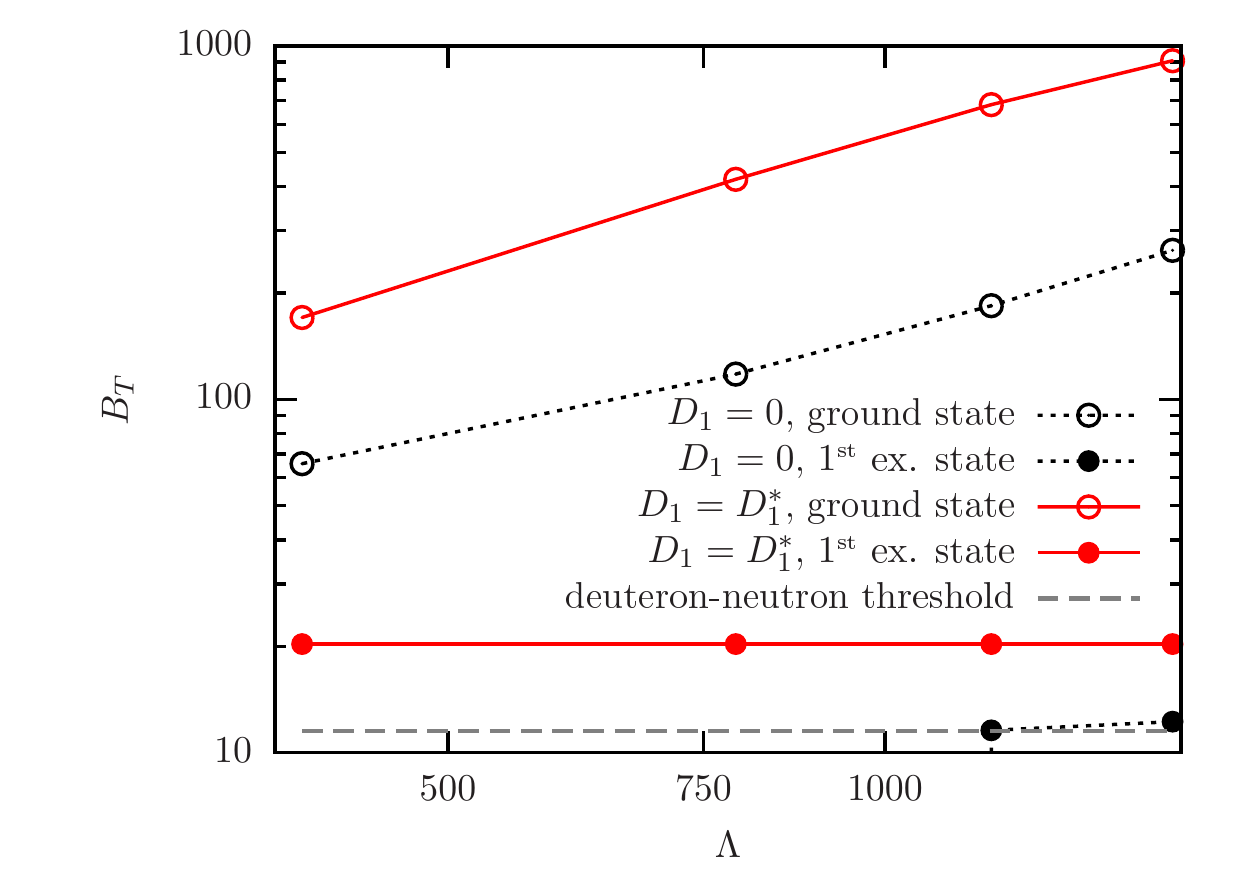}
\caption{\label{fig_tspec} Dependence on the cutoff $\Lambda$ (in MeV)
of the three-nucleon bound-state 
spectrum $\eotri$ (in MeV) in the triton channel at \mpi=510~MeV.
The neutron-deuteron threshold is indicated by a dashed line.
Empty (filled) circles mark ground (excited) bound states. 
The dotted lines are for $D_1=0$.
For the full lines the three-nucleon force 
fixes the shallow state to $\eotri$.}
\end{center}
\end{figure}

We can also gain some insight into the cutoff dependence of $D_1$.
In the absence of a three-nucleon force, the triton
spectrum depends sensitively on $\Lambda$, indicating a lack of
renormalizability. 
The example of $m_\pi=510$ MeV is shown in Fig.~\ref{fig_tspec}.
When $D_1=0$, the open circles on the dotted line show an almost exponential
dependence of the ground state on the cutoff.
As indicated by the filled circles on another dotted line,
around 1.2~GeV 
a second bound-state pole emerges, which also becomes increasingly
more bound as the cutoff increases. The pattern repeats as the
cutoff increases further.
Renormalizability can be achieved with the non-derivative three-body force
\cite{Bedaque:1998kg,Bedaque:1998km,Bedaque:1999ve}
\begin{equation}\label{eq_d1_coord}
\mn \Lambda^{-2} D_1 (\Lambda)= F(\Lambda/\Lambda_*)  \,,
\end{equation}
where $\Lambda_*$ is an \mpi-dependent parameter that determines the three-body
spectrum and $F$ is a dimensionless function that depends on the regulator and
on which state is kept at the observed $\eotri$.
Accordingly, in the right panel of Fig.~\ref{fig_lec}
we display $\mn \Lambda^{-2} D_1$.


For all values of \mpi~we fit $D_1$ to ensure the deep bound state
remains at the observed value of $\eotri$,
in which case $F$ increases monotonically with $\Lambda$.
The resultant values of the LEC define bands which vary significantly
in width with \mpi.
For \mpi=805~MeV (blue band around pentagons in the right panel 
of Fig.~\ref{fig_lec}) 
and the physical \mpi~(black squares), the band width
is narrow relative to \mpi=510~MeV (gray band around circles). 
All three bands correspond to repulsive interactions. 
This means that
without a three-nucleon force there is a three-body state which is 
more bound than the observed triton.
This state is then ``raised'' to the triton by the repulsive interaction. 

Since additional three-body bound-state poles appear at the two-body 
threshold with increasing cutoff,		
it is 
possible to renormalize $D_1$ to 
a shallow state instead.
In this case, $F$ changes.
In the example of Fig.~\ref{fig_tspec}, we can
fit $D_1(\Lambda)$ so that the first excited state 
is ``lowered'' to the triton level 
as indicated by the filled circles on a full line
--- we label the corresponding values of $D_1$ as $D_1^{*}$
in Table~\ref{tbl:LECs} and Fig.~\ref{fig_tspec}.
In this case, two states remain bound: the triton and a deep state 
shown by the empty circles on full line in Fig.~\ref{fig_tspec},
the latter with a binding energy that goes from 170~MeV at $\Lambda=400~$MeV to 
900~MeV at $\Lambda=1.6~$GeV.
The increased binding of the deep state compared to its binding when $D_1=0$
shows that the force is attractive in this range of cutoffs, and it indeed
has an opposite sign to the force that keeps the
ground state at $\eotri$, as seen in Fig.~\ref{fig_tspec}.

The functional dependence of the three-nucleon LEC in Fig.~\ref{fig_lec} is,
by construction, identical
to that of the binding energies on the cutoff when $D_1=0$.
The latter grows faster than quadratic
(upward bending of the black dotted line with empty circles 
in Fig.~\ref{fig_tspec}).
This deviation is consistent with the increase of $D_1$ 
in Fig.~\ref{fig_lec}
which is not just quadratic but receives contributions from higher powers 
of $\Lambda$.
For both fitting choices,
we find the uncertainty in $D_1$ by taking $\eotri\in[15.8,24.8]~$MeV 
(see Table~\ref{tab.scales}) at \mpi=510~MeV.
It is considerably larger when a repulsive three-nucleon force is used,
as shown by the width of the gray band in the right panel 
of Fig.~\ref{fig_lec},
which is much larger than the red band that represents
the variation in the attractive force strength.
In contrast to the log-periodic behavior of the three-body force as a 
function of the cutoff found in Refs. 
\cite{Bedaque:1998kg,Bedaque:1998km,Bedaque:1999ve}
we find both, the central value and the uncertainty, 
to increase monotonically with $\Lambda$
for all \mpi~except for the calibration to the
excited state (empty circles in Fig.~\ref{fig_lec}). 
No discontinuities at critical values of $\Lambda$ are observed because the
eigenstate we chose to fit $D_1$ was always either the ground or the first 
excited state. A log-periodic $F$, as in Refs. 
\cite{Bedaque:1998kg,Bedaque:1998km,Bedaque:1999ve},
is found if the LEC is calibrated always to the shallowest state in the 
spectrum. 
As a consequence, after renormalization the smallest binding energy
is fixed, while
states accrete from very large binding energies at the critical
values of $\Lambda$. 

The 
significant difference in uncertainty of the three-nucleon-force parameter
when fitted 
to the ground or excited state is related to the functional dependence of those
states on $\Lambda$.
In the vicinity of a critical $\Lambda$ where an additional state 
enters the spectrum, the eigenvalue of the excited state increases much 
slower than that of the ground state
(compare slopes of the dotted lines in Fig.~\ref{fig_tspec}).
The respective three-nucleon interaction strength inherits a larger slope 
if the ground state is fitted
relative to calibrating the excited state. 
Since
both input and regulator variation represent a change in the renormalization
scheme, 
the observed difference in uncertainty is a consequence of 
the differences in slope.



Different values of $\Lambda$ and different regularization schemes
correspond to different models of the short-distance behavior of the theory. 
These models might allow for deeply bound states in the deuteron, triton, and 
alpha-particle spectra. 
A tenet of EFT is that high-energy, or short-distance, phenomena 
can be accounted at each order by the most general interactions
consistent with symmetries and required by RG invariance.
In the specific case, we use this tenet to conjecture that
low-energy observables, such as the nD scattering lengths,
should be independent of whether we fit the triton
to the deepest, second-deepest, ..., or shallowest state.
This has been seen in simple explicit examples, 
such as Ref. \cite{NakaichiMaeda:1995zz},
where invariants of few-body spectra
were analyzed with respect to changes in the short-distance structure of the 
employed models.
It is not the scope of this work to assess differences between the 
various schemes, and hence we employ repulsive three-nucleon forces consistently
in all calculations below.

\section{Results}
\label{sec_res}

The Phillips \cite{Phillips:1968zze} and Tjon \cite{Tjon:1975} correlations are 
non-trivial features of nuclear physics.
Their sensitivity with respect to \mpi~is analyzed
here. In addition, we consider the quartet three-nucleon channel which is 
less sensitive to the short-distance structure of the interaction, \ie,
no three-nucleon interaction contributes up to high order.
With these predictions, we conclude that key nuclear properties are,
qualitatively,
insensitive to \mpi~---~a conjecture based on the universal EFT approach. 
We compare the results for \mpi=510~MeV and \mpi=805~MeV with 
LO \eftnopi~results
at the physical pion mass to make similarities and differences explicit.

Besides identifying 
the peculiarities 
of large pion masses, we 
predict the outcomes of
``experiments'' in these hypothetical worlds.
As described in the previous section,
for each cutoff, \ie, model for unobservable short-distance structure, 
\eftpihi~differs in its coupling constants.
If these models differ in predictions by a finite amount, 
it is this amount that quantifies the theoretical error.
In the physical world, theoretical error estimates of this kind were used 
previously to 
make predictions through 
correlations (see {\it e.g.} Refs. \cite{Kirscher:2011zn,Hammer:2014rba}).
Since the theoretical error for $\eotri$ and $\eotet$ is large relative to 
the experimental one,
the Phillips and Tjon lines at LO in \eftnopi~do not
constrain observables
further at physical \mpi.
At larger \mpi, however, the lattice uncertainty is still significant 
(see Table~\ref{tab.scales}) and,
\textit{a priori}, there is no reason why \eftpihi~should not be able to 
constrain observables
more tightly
through those correlations than solely by the ``experimental'' error.

We assess sensitivity of results to higher-order terms in the 
EFT expansion 
by a variation of the cutoff-regulator parameter 
in the range $\Lambda\in\lbrace 2, 4, 6, 8\rbrace~$fm$^{-1}$.
This range includes the critical value for appearance at the physical pion mass
of an excited state, when $D_1=0$
(see discussion of Fig.~\ref{fig_tspec} in Sec.~\ref{sec_calib}).
For lattice pion masses, the upper limit is 2-3 times the expected
breakdown scale, where in general we see signs of convergence
in observables.
Although we might ideally want even higher cutoffs at the expense
of further computational time,
our estimate of the truncation error is probably not an underestimate 
because we include cutoff values below the expected breakdown scale.
Such low cutoffs introduce larger errors than the truncation.
A more reliable error estimate has to await higher-order calculations
where the breakdown scale manifests itself.

\subsection{The Three-Body Sector}\label{sec_3}

The physical nucleon-deuteron system splits into two significantly different 
spin channels: doublet (or triton) with $s=1/2$
and quartet with $s=3/2$. 
The former (latter) supports (does not support) a bound state. 
In the doublet channel, an additional counterterm
enters at LO --- $D_1$ term in Eq.~\eqref{hpsi}
--- 
while the quartet channel is renormalized
with $C_{S,T}$, only. 
The consequences to 
large \mpi~are the subject of the following two sections.
Since we include no Coulomb interactions, our results 
at physical pion mass apply only to neutron-deuteron scattering.

\subsubsection{Neutron-deuteron 
$^4S_{3/2}$ channel}
\label{quartet_section}

\begin{figure}[t]\begin{center}
\includegraphics[width=0.5 \textwidth]{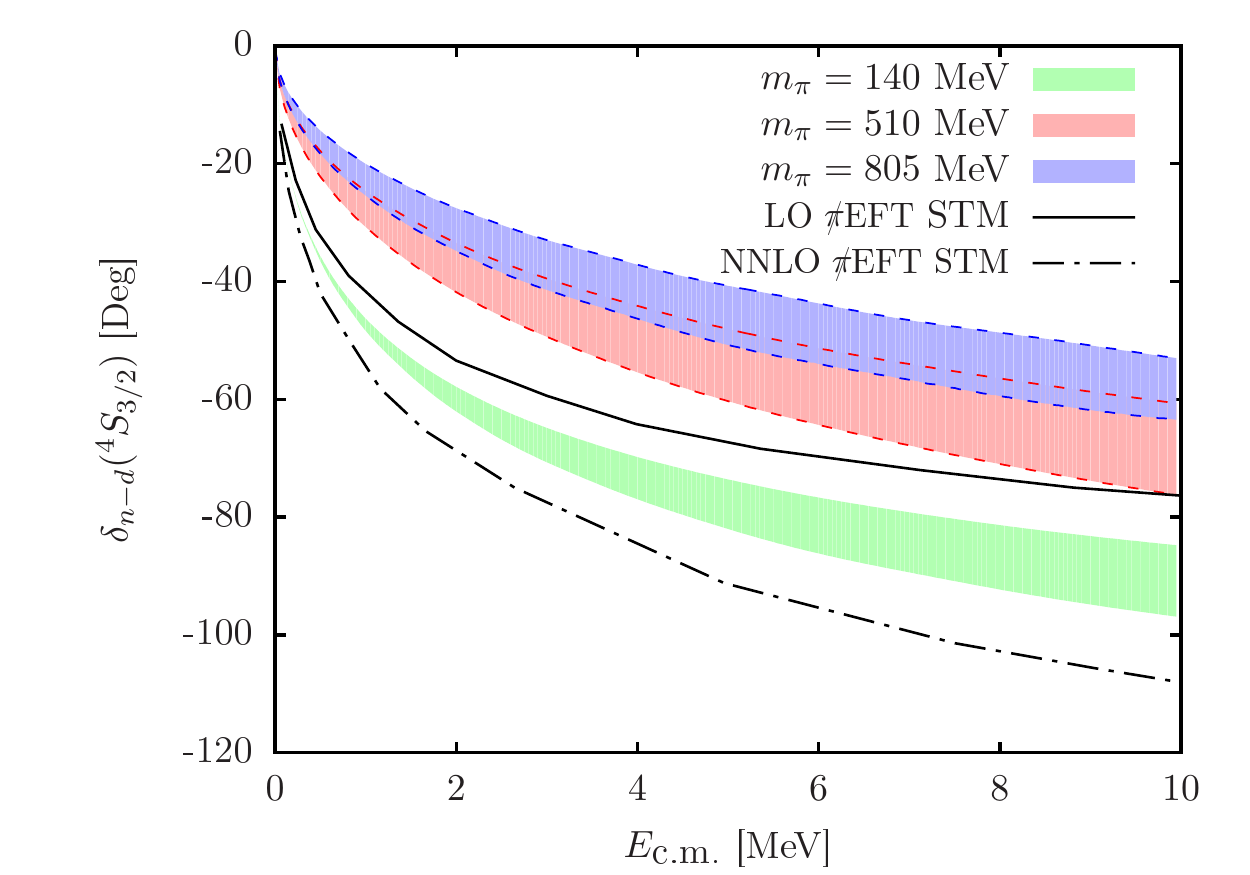}
\caption{\label{fig_quartet_phases} 
Elastic neutron-deuteron scattering phase shift $\delta$ in the spin-quartet 
$^4S_{3/2}$ channel (in degrees) for various pion masses
as function of the center-of-mass energy $E_{c.m.}$ (in MeV). 
The green (\mpi=140~MeV), red (\mpi=510~MeV), and blue (\mpi=805~MeV) 
shaded areas are the LO \eftpihi~results of the \rgm~for a cutoff in 
the range $[2,8]~\text{fm}^{-1}$. 
For the overlapping $m_\pi=510,805~$MeV uncertainty bands, the upper (lower) 
edge, corresponding to $\Lambda=8 \, (2)~$fm$^{-1}$, is given by dashed lines. 
For \mpi=140~MeV, the solid (dashed-dotted) black line represents 
LO (N$^2$LO) \eftnopi~results 
from the solution of the STM equation \cite{Bedaque:1998km,Bedaque:1999vb}.}
\end{center}
\end{figure}

The phase shifts in the quartet channel can be calculated
in LO solely on the basis of two-nucleon input.
In Fig.~\ref{fig_quartet_phases}, we show our phase shift results for elastic 
nD scattering below $10~$MeV, calculated with the \rgm. 
For all three pion masses, the cutoff variation
between 2~fm$^{-1}$ and 8~fm$^{-1}$ is shown by green (\mpi=140~MeV), 
red (\mpi=510~MeV), and blue (\mpi=805~MeV) bands. 
The upper (lower) edge corresponds to 8 (2)~fm$^{-1}$. 
For the physical \mpi, we compare our results to previous 
LO and N$^2$LO \eftnopi~calculations~\cite{Bedaque:1998km,Bedaque:1999vb} 
(black solid and dashed-dotted lines)
obtained from the solution of the 
Skorniakov--Ter-Martirosian (STM) equations.
The difference between these curves is, of course, 
a good reflection of the uncertainty of the LO calculation
at the physical pion mass.
Our result has the correct energy dependence and
lies between the two curves.
Our error band 
accounts for cutoff variation but not numerical uncertainty. 
The latter is
included in the postdiction of the nD scattering length given below.
The energy dependence and band widths
are similar
for the three values of the pion mass we consider.
This suggests an invariance with respect to \mpi~of the uncertainty --- 
and therefore the
convergence rate of the EFT.

\begin{widetext}
\begin{figure*}[tb]\begin{center}
\includegraphics[width=1. \textwidth]{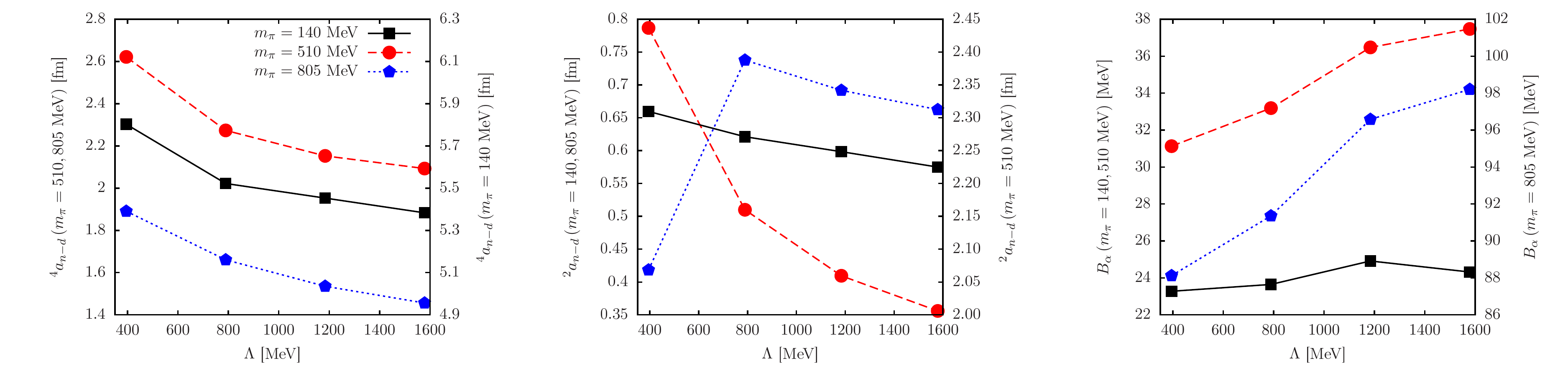}
\caption{\label{fig_cutoffdep}
Dependence on the cutoff $\Lambda$ (in MeV)
of the quartet ($\andq$, left) and doublet ($\andd$, center) neutron-deuteron 
scattering lengths (in fm), and of the
alpha-particle binding energy $\eotet$ (right, in MeV), for 
$m_\pi=140$ MeV (black squares), 
$510~$MeV (red circles), and
$805~$MeV (blue pentagons).
}
\end{center}
\end{figure*}
\end{widetext}

We extract a scattering length at $\ecm=0.001~$MeV. 
The cutoff dependence is illustrated in the left panel
of Fig.~\ref{fig_cutoffdep}. For all values of the pion mass
we observe a nice convergence pattern.
Our final values are shown in the first row of Table~\ref{tab_res_summ}.
The errors are the sum of the sensitivity to higher-order effects assessed with 
the cutoff variation, and
the numerical uncertainty, which we measured to be less than $1~$fm.
They are of similar size for the three pion masses,
as for the phase shifts at higher energies.

\begin{table*}
\renewcommand{\arraystretch}{1.1}
\caption{\label{tab_res_summ}{\small Leading-order 
postdictions (\eftnopi) and predictions (\eftpihi) 
for the quartet and doublet neutron-deuteron scattering lengths
$\andq$ and $\andd$
at three pion masses, in comparison with experiment and LQCD.
The theoretical uncertainty considers cutoff variation between 2~fm$^{-1}$ 
and 8~fm$^{-1}$, model-space truncation, and LQCD-input variation.}}
\begin{tabular}{rccc}
\hline\hline
&\multicolumn{1}{c}{\eftnopi}&\multicolumn{2}{c}{\eftpihi} \\
\hline
\mpi~[MeV]         & $140$          & $510$          & $805$        \\ 
\hline
$\andq~$[fm]       &  $5.5\pm 1.3$  & $2.3\pm 1.3$   & $1.6\pm 1.3$ \\
$\andd~$[fm]       &  $0.61\pm 0.50$ & $2.2\pm 2.1$ & $0.62\pm 1.0$ \\
\hline
& \multicolumn{1}{c}{experiment \cite{Dilg:1971}}&\multicolumn{2}{c}{LQCD} \\
\hline
$\andq~$[fm]       &  $6.4\pm 0.020$  & ? & ?  \\
$\andd~$[fm]       &  $0.65\pm 0.040$ & ? & ?  \\
\hline\hline
\end{tabular}
\end{table*}



The quartet scattering length is an example of what is sometimes
called a low-energy theorem: to a high order it is entirely determined
by LECs fixed in other processes.
The value we obtain for $\andq$ at physical pion mass is 
consistent with the \eftnopi~postdictions \cite{Bedaque:1997qi}
of $\andq=5.1\pm 0.80$ at LO 
and $\andq=6.4\pm 0.020$ at N$^2$LO,
and with the experimental value \cite{Dilg:1971}.
We find a slow decrease with \mpi, but no significant change, 
which could have arisen if there were a shallow bound state in this channel.
The ERE should apply
below the deuteron break-up threshold, where the deuteron can be treated
as a single body. We might expect that, barring some fine-tuning,
the size of the ERE parameters is set by 
the deuteron break-up threshold.
$k_{\rm pn}\simeq \sqrt{4\mn\eodim/3}$.
The numbers in Table \ref{tab_res_summ} show indeed very good agreement
with the expectation $|\andq|={\cal O}(1/k_{\rm pn})$.

Both the convergence pattern with the cutoff (reflected in 
error bands) and the natural size of the resulting observables
suggest that \eftpihi~behaves in similar ways to \eftnopi.
There is no evidence that observables in this channel require a 
different treatment
from an EFT point of view for the larger pion masses, 
\ie, the same power counting is applicable.

\subsubsection{Neutron-deuteron 
$^2S_{1/2}$ channel} 
\label{doublet_section}

As precise calculations of the three-nucleon system became possible
in the late 1960s, correlations were observed among certain three-body
observables calculated with a variety of potentials fitted to two-nucleon data.
The best-known example is the Phillips correlation \cite{Phillips:1968zze}
between the doublet scattering length $\andd$~and the triton
binding energy $\eotri$.
In \eftnopi~these correlations are understood 
\cite{Bedaque:1998kg,Bedaque:1998km}
by the fact that,
if the two-body input is fixed, three-body predictions in the
doublet channel still depend on one parameter in LO,
which determines the three-body force in
Eq. \eqref{eq_lo-lagrangean}. As this one parameter is varied,
three-body observables sensitive to the LO three-body force 
all change in a correlated way. 
$\eftpihi$, by construction, predicts the correlations, and
we establish here similar higher-order uncertainties for all
pion masses.

Fixing the LO three-body parameter to one datum,
other observables are calculated as for the quartet channel.
The cutoff dependence of the doublet neutron-deuteron
scattering length is shown in the middle panel
of Fig.~\ref{fig_cutoffdep}
in the case where the three-body parameter is determined by $\eotri$,
as described in Sec. \ref{sec_calib}. 
Again signs of convergence are visible, but
not as clearly as for the quartet scattering length.
The lowest cutoff of 2 fm$^{-1}$ is not clearly above the
breakdown scale at $m_\pi=805$ MeV, 
and indeed it generates significant errors.
Therefore, for this pion mass we consider only cutoffs 4 fm or higher
in the following analysis.

One also expects the values of the doublet scattering length
to be correlated with the triton binding energy.
This is particularly clear when 
$|\eotri-\eodim|\ll \eodim$, 
in which case the triton can be described as a neutron-deuteron bound state
and the small binding translates into $1/\andd\ll 1$.
But this correlation is observed also beyond this region:
in Fig.~\ref{fig_phillips}, 
the Phillips correlation is shown for the three pion masses.
As the renormalization condition fixes the binding energy
but does not eliminate a residual cutoff dependence, 
which can only be removed by higher-order interactions,
the correlation is manifest as a band of finite width rather than a 
one-dimensional line.
This width represents the theoretical error at LO \eftpihi. 
The bands were mapped out by a line
for each cutoff $\Lambda\in\lbrace 2,4,6,8\rbrace~\text{fm}^{-1}$. 
Each such line is
parametrized by a factor multiplying the three-body interaction.
At \mpi=805~MeV, the correlation is about to break down for the lowest 
cutoff, $\Lambda=2~\ifm$ (blue
dashed line in Fig.~\ref{fig_phillips}),
which is another evidence that this value cannot be considered
representative of the EFT truncation error.

\begin{figure*}[tb]\begin{center}
\includegraphics[width=1.0 \textwidth]{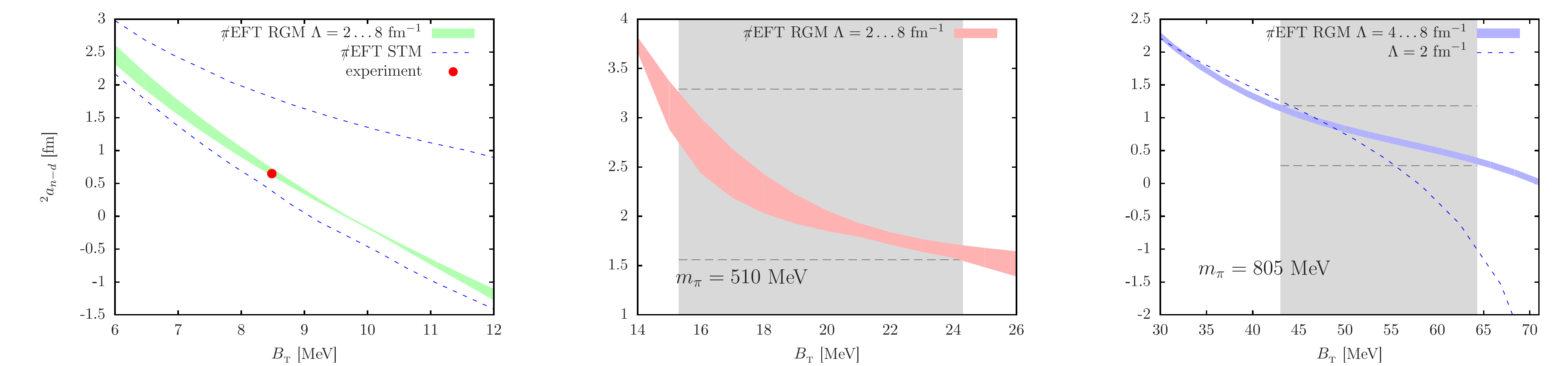}
\caption{\label{fig_phillips} Correlation between 
the doublet neutron-deuteron scattering length $\andd$ (in fm)
and the three-nucleon binding energy $\eotri$ (in MeV). 
The green (\mpi=140~MeV, left panel) and red (\mpi=510~MeV, center)
shaded areas are the LO \eftpihi~results 
of the \rgm~for a cutoff in the interval $[2,8]~\text{fm}^{-1}$. 
The blue (\mpi=805~MeV, right panel) shaded area is the LO \eftpihi~results 
of the \rgm~for a cutoff in the interval $[4,8]~\text{fm}^{-1}$,
with the cutoff $\Lambda=2$ fm$^{-1}$ shown as a blue dashed line.
For \mpi=140~MeV (left panel), experimental data are marked with a red dot 
and blue dashed lines display LO \eftnopi~results 
obtained with the STM equation for sharp cutoffs of 140 and 900 MeV
\cite{Griesshammer:2004pe}. 
The gray shaded area (center and right panels) marks the
lattice uncertainty in $\eotri$.
Values for $\andd$ between the horizontal dashed lines are consistent with all 
other 
low-energy data.}
\end{center}
\end{figure*}

We extract the values for $\andd$ shown in the 
second row of Table~\ref{tab_res_summ}.
In the doublet channel, a too-small model space can be more
easily detected than in the quartet channel
from an under- or over-bound triton.
As a consequence, the numerical \rgm~uncertainty 
is about $0.1~$fm and therefore small relative 
to higher-order effects which are taken as the width of the band: 
$0.26~$fm for \mpi=510~MeV, and
$0.13~$fm for \mpi=805~MeV.
The approximately constant width of the Phillips band for all three $m_\pi$
suggests invariant higher-order uncertainty with increasing $\eotri$.
Since higher-order effects scale with momenta --- 
those of nucleons increase in the triton as $\eotri$ increases --- 
the band should intuitively widen towards larger $\eotri$.
In effect, lattice input uncertainty dominates the theoretical error.
For the two unphysical pion masses (center and right panels), 
the gray-shaded areas represent
data uncertainty given in Table~\ref{tab.scales}.
The intersections of the edges of the error bands with the correlation bands 
define areas (gray areas bounded by horizontal dashed lines) in the 
$\eotri-\andd$ plane which contains pairs of values
that are consistent with all other data points.
The total theoretical uncertainty as estimated in Table.\ref{tab_res_summ}
includes the error in
the LQCD input,
$1.71~$fm for $m_\pi=510~$ MeV and $0.26~$fm for $m_\pi=805~$MeV. 

The calculation at physical \mpi~(left panel in Fig.~\ref{fig_phillips})
serves as a benchmark. 
The dashed lines represent the  
solution of the STM equation at
LO in $\eftnopi$~when a sharp cutoff was varied between from 140 to 900 MeV
\cite{Griesshammer:2004pe}. 
Experiment is represented by the red dot.
Our band is consistent with both.
The small scattering length compared to the break-up threshold inverse 
momentum,
$1/k_{\rm pn}\simeq 2.2$ fm,
is a sign of a zero
of the $T$ matrix near the nD threshold.
As discussed in Ref. \cite{vanKolck:1998bw},
such a zero is located at 
$k_0^2\sim -\andd \, k_{\rm pn}^{3}$ 
with respect to the origin of the complex relative-momentum plane,
when $|k_0|\ll k_{\rm pn}$ and the deuteron can be treated
as elementary.
A consequence is a large effective range 
$^2r_{\rm nD}\sim -(\andd \, k_0^2)^{-1}$
and a small radius of convergence of the usual ERE.
Data suggests $|k_0|\sim 20$ MeV on the imaginary momentum axis, and
indeed this is what was found by explicit calculation already many years ago
\cite{Whiting:1976zz}.
The negative slope of the Phillips line indicates that, as the 
three-body force is changed so that the
triton gets more bound, this pole crosses threshold.
The zero remains in the region of validity of the elementary-deuteron 
theory for 1 or 2 MeV around the physical value of $\eotri$.
In this region a modified ERE \cite{vanOers1987} holds \cite{vanKolck:1998bw}.


In the center and right panels of Fig.~\ref{fig_phillips})
we extend LQCD to the realm of few-body scattering,
which is not as easily accessible directly on the lattice.
The negative slope of the Phillips line persists but
the line moves up in the $\eotri-\andd$ plane, 
and it gets flatter, as $m_\pi$ increases.
The ``measured'' triton binding energy rises monotonously 
(Table \ref{tab.scales}),
the doublet scattering length first increases then decreases
(Table \ref{tab_res_summ}).
For $m_\pi=510, 805$ MeV, the accidental zero 
of the nD scattering amplitude is no longer clearly present
in the region where the deuteron can be taken as elementary,
and no particularly large effective range is expected.

The increasing~$\andd$ with increasing $\eotri$
from 140~MeV to 510~MeV pion mass is opposite to the trend
found for fixed pion mass, as identified above.
It is not the triton binding energy but the triton-to-deuteron 
binding ratio shown in
Table~\ref{tab.scales} which decreases with increasing $\andd$.
This ratio is important because it measures the motion
of the ``experimental'' point in the $\eotri-\andd$ plane:
$\eodim$ influences (together with $\eonn$) the position of the line, and
$\eotri$ fixes the position along the line.
As the ratio $\eotri/\eodim$
decreases from \mpi=140~MeV to \mpi=510~MeV and increases 
from 510~MeV to 805~MeV pion mass, the three-nucleon bound state 
comes closer to and farther away from the nD threshold.

In particular, the relatively large scattering length at $m_\pi=510$ MeV
reflects a less-bound triton relative to
the neutron-deuteron threshold.
Once the errors are considered,
$\eotri$ is just a few of MeV away from $\eodim$ at $m_\pi =510$ MeV.
In fact, in contrast to $m_\pi =140, 805$ MeV, the binding momentum of the last
nucleon can be smaller than the deuteron break-up momentum,
$\kappa_{\rm nD}\simeq \sqrt{4\mn(\eotri-\eodim)/3}< k_{\rm pn}$,
in which case the ERE is likely to apply. This leads
to the prediction
\begin{equation}
\andd= \frac{1}{\kappa_{\rm nD}}
\left( 1+ \frac{^2r_{\rm nD}\, \kappa_{\rm nD}}{2}+\ldots \right).
\label{anddERE}
\end{equation}
The first term gives, for the central values of the binding energies,
$\andd=1.6$ fm with a correction of about 50\% from the second
term if $|^2r_{\rm nD}|\sim 1/k_{\rm pn}$. This estimate agrees with the 
central value calculated
with the full three-body dynamics given in Table \ref{tab_res_summ}.
Barring significant shifts in the central values as lattice
errors are reduced, in this lattice world
the triton can be viewed as a two-body halo system.
\par
It is an open question whether there is a pion mass, 
possibly around 510 MeV, where
the triton converges to the deuteron threshold. 
If there is, we would be witnessing 
a qualitatively new phenomenon in few-nucleon
physics.
The Efimov effect in the three-body system 
is a prominent example of universal feature emergent from
the unitary limit in the two-body sector. 
A pion mass which produces the analog three-body unitarity, 
$1/\andd\to 0$,
would 
be a world where the four-body system exhibits an
Efimov-type spectrum.


\subsection{The Four-Body Sector}

While there is no lattice data on three-nucleon scattering observables and 
thus the results presented
in the previous subsection remain to be verified ``experimentally'',
 \ie, with a direct LQCD calculation,
there is data on the ground-state energy of the four-nucleon system.
In this section,
we find the three- and four-nucleon ground-state energies correlated for all 
three \mpi.
At the physical \mpi, the relation is known as the Tjon line \cite{Tjon:1975}
which can again be explained by a variation in the single
LO three-body force parameter.

In Fig.~\ref{fig_tjon}, the 
correlations between the ground-state energies 
of the three- and four-nucleon systems
are shown.
The different graphs represent results for the three pion masses.
We observe an increase in alpha-particle binding in step with the increase
in triton binding energy, which 
is not surprising in pionless EFT
because
with fixed two-nucleon input 
it is the same three-body force that controls the binding
of the three- and four-nucleon systems.
The correlation is  manifest in a band, not a line, and the width of 
the band measures the
theoretical uncertainty assessed via cutoff variation.
With the central value
of $\eotri$ as input in the three-body force, 
the dependence on the cutoff
of the alpha-particle binding energy $\eotet$
is shown on the right panel of Fig.~\ref{fig_cutoffdep}.
The slope of the correlation
\textit{lines} --- as before, each line is parametrized by a variation of 
the $D_1$ three-body interaction
strength --- further affects the LO \eftpihi~uncertainty.
The larger the slope, the larger the uncertainty in $\eotet$ 
due to the uncertainty in $\eotri$.

\begin{figure*}[tb]\begin{center}
\includegraphics[width=1. \textwidth]{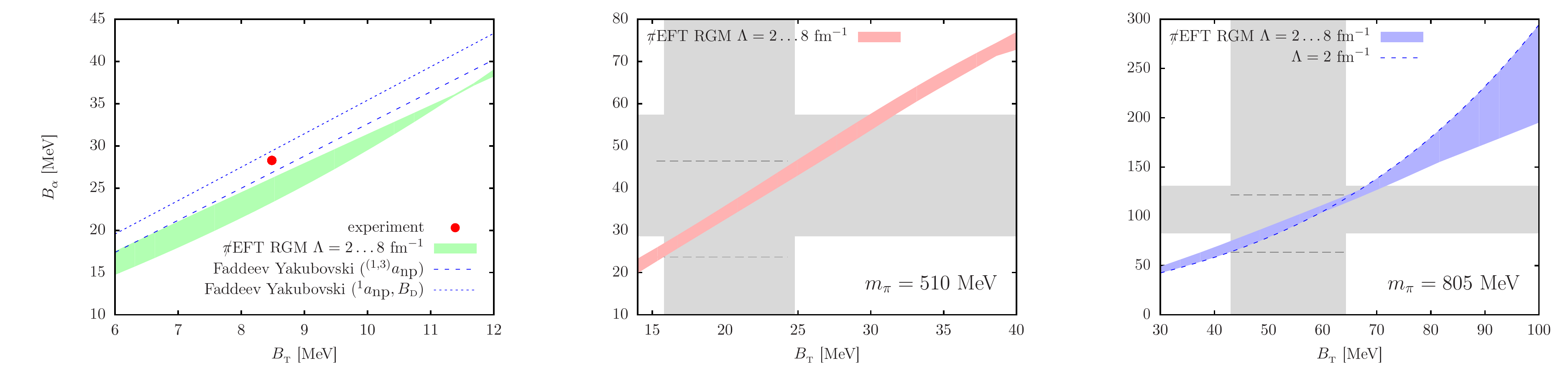}
\caption{\label{fig_tjon} Correlation between the three- ($\eotri$,
in MeV) and four- ($\eotet$, in MeV) nucleon binding energies.
The green (\mpi=140~MeV, left panel), 
red (\mpi=510~MeV, center panel) and blue (\mpi=805~MeV, right)
shaded areas are the \rgm-LO~\eftpihi~results for a cutoff 
in the interval $[2,8]~\text{fm}^{-1}$. 
For \mpi=140~MeV (left panel), the green uncertainty band
represents sensitivity to the cutoff \textit{and}
to the renormalization input (whether $\eodim$ or $\anpt$).
Experimental data are marked with a red dot, 
and the blue dotted (dashed) line represents LO \eftnopi~results
from Ref. \cite{Platter:2004zs}
using $\anps,\anpt$ ($\anps,\eodim$) as input. 
The gray shaded areas in the center 
and right panels mark lattice uncertainty in $\eotri$ and $\eotet$.
Values for $\eotet$ between the horizontal dashed lines are consistent 
with all other low-energy data.
}
\end{center}
\end{figure*}

For the physical pion mass, our error band 
does not agree well with the LO results of Ref. \cite{Platter:2004zs}
shown in the left panel of Fig.~\ref{fig_tjon}.
In Ref. \cite{Platter:2004zs}
the alpha-particle binding energy was found by
a solution of the Faddeev-Yakubovski integral equations 
with a Gaussian 
regulator on the relative incoming ($p'$) and outgoing ($p$)
momenta, $\exp[-(p^2+p'^2)/\Lambda^2]$.
The 
uncertainty was assessed by a cutoff variation
$\Lambda\in[8,10]~\textrm{fm}^{-1}$,
thus excluding
a reported stronger cutoff
dependence for $\Lambda<8~\textrm{fm}^{-1}$.
The cutoff variation was then deemed small compared to
the higher-order uncertainty estimated by 
changing the two-body input:
the two curves obtained from $\eodim$ and $\anpt$ 
are represented by the blue lines in Fig.~\ref{fig_tjon}.
We have similarly examined the input dependence:
for
$\Lambda=8~$fm$^{-1}$, we replaced $\eodim$ with $\anpt$ and
found $\eotet$ 
(upper bound of the correlation band for \mpi=140~MeV,
green area, left panel in Fig.~\ref{fig_tjon}) 
larger by $2~$MeV.
Even with this extended variation of the renormalization scheme,
the two uncertainty bands do not overlap.
In contrast, the current \rgm~results for the Tjon correlation
band are consistent with the previous \rgm~LO-\eftnopi~calculation
of Ref. \cite{Kirscher:2009aj}.
The convergence of $\eotet$ to the physical value when the NLO potential is 
iterated \cite{Kirscher:2009aj}
suggests that in both LO calculations the theoretical error as shown by the 
band widths in Fig.~\ref{fig_tjon} is a lower bound.
For our theoretical error estimates, we interpret \rgm~and Faddeev 
calculations, \ie, different regularizations and model-space cutoffs, 
as different renormalization schemes. For $\eotet$ and physical \mpi, 
the uncertainty is thus given by the
spread of results of both methods 
(difference between short-dashed blue line and lower edge of the green band). 

For unphysical pion masses we indicate,
as before, the uncertainty in LQCD energies
by gray bands in the center and right panels
of Fig.~\ref{fig_tjon}:
a vertical band for $\eotri$ and a horizontal band for $\eotet$.
Values for $\eotet$ in an interval bounded by the intersection of the upper 
(lower) edge of the
\eftpihi~uncertainty band with the right (left) boundary of the band of 
LQCD-allowed $\eotri$ values
are indicated by horizontal dashed lines.
This range is slightly larger than the constraint already given by 
``experiment'' for \mpi=805~MeV, and slightly narrower for \mpi=510~MeV.
However, given the renormalization-input dependence seen at physical
pion mass, we estimate the theoretical uncertainty by
conservatively doubling the width of the \rgm~correlation band,
plus 2~MeV as an upper bound of the numerical uncertainty 
(see Fig.~\ref{fig_rgm_eihh} for this estimate),
plus the experimental LQCD uncertainties in $\eotet$.

The results for the alpha-particle binding energy
are summarized in Table~\ref{tab_res_4}. 
The predicted value is taken 
as the central value in the uncertainty band. 
The EFT results---absolute binding energies and the ratios---are consistent with experiment at physical pion mass
and with LQCD at higher masses given the uncertainty estimates on both,
experimental and theoretical side.
One should keep in mind that the experimental number reflects
the additional effects of the Coulomb interaction between protons,
which does not enter the LQCD results. 
As discussed in Sect. \ref{sec_PEFT}, Coulomb effects should be of higher order
in the relatively tight helion and alpha-particle ground states.

\begin{table*}
\renewcommand{\arraystretch}{1.1}
\caption{\label{tab_res_4}{\small Predictions for the four-nucleon 
binding energy $\eotet$ and the universal alpha-to-triton ratio $\eotet/\eotri$
from LO pionless EFT at three pion masses,
in comparison with experiment and LQCD. 
The theoretical errors 
include numerical and EFT uncertainty. The uncertainty in the fractions
(lines 2 and 4) adds independent errors in quadrature.}}
\begin{tabular}{cccc}
\hline\hline
\mpi~[MeV] & $140$       & $510$     & $805$        \\ 
\hline
&\multicolumn{1}{c}{\eftnopi}&\multicolumn{2}{c}{\eftpihi} \\
\hline
$\eotet~$[MeV]       & $24.9\pm 4.3$ & $35\pm 22$ & $94\pm 45$\\
$\eotet/\eotri$      & $2.9\pm 0.51$         & $1.7\pm 1.1$          & $1.8\pm 0.9$ \\
\hline
& \multicolumn{1}{c}{experiment}&\multicolumn{2}{c}{LQCD} \\
\hline
$\eotet~$[MeV]       & $28.3$ & $43.0\pm 14.4$ & $107.0\pm 24.2$ \\
$\eotet/\eotri$      & $3.34$ & $2.1\pm 0.85$          & $2.0\pm 0.6$ \\
\hline\hline
    \end{tabular}
\end{table*}

As expected, the $\eotet-\eotri$ correlation has a positive slope
for any pion mass.
For each correlation band, we define the slope with a linear regression 
through the $\eotri,\eotet$ pairs predicted with \eftpihi~for
all cutoff values which are within data uncertainty (gray areas) only.
At physical \mpi, our calculation yields a smaller slope ($\approx 3.6$) than 
Ref. \cite{Platter:2004zs} ($\approx 3.8$).
With increasing \mpi, the slope is found to decrease, 
$\approx 2.1$ for \mpi=510~MeV and \mpi=805~MeV.
In other words, the ratio $\eotet/\eotri$
\textit{does} change with \mpi, 
consistent with the lattice extractions, as shown in Table~\ref{tab_res_4}.

Since in obtaining the Tjon line the three-body
force is being varied, 
the structure of the line (slope, curvature, intercept) must depend
on the two-nucleon interactions.
Indeed, from the ratios listed in Table~\ref{tab_res_4} we infer that whatever
leads to the different ratios between the 
deeper two-nucleon state --- recall that for unphysical masses, the interaction
in the $^1S_0$ channel sustains a bound state, see Table~\ref{tab.scales} ---
and the triton is not the main factor behind the change in the
slope of the Tjon line.
As for the Phillips line,
the structure of the Tjon line depends on both pieces of two-nucleon input.
For example,
in Fig.~\ref{fig_tjon_slope}, we demonstrate the sensitivity of the slope of 
the Tjon line with
respect to the pole in the spin-singlet two-nucleon amplitude. 
The triplet binding energy, \ie,
the deuteron was held fixed to the lattice central
value at \mpi=510~MeV , $\eodim=11.5~$MeV.
Three cases are shown for 
$\Lambda=4~\text{fm}^{-1}$, corresponding to different 
calibrations of $C_{0,1}$, the LEC controlling the channel:
{\it i)} 
a singlet neutron-proton state with binding energy of approximately 
$11.5~$MeV, \ie, the deuteron energy;
{\it ii)} 
a shallow bound singlet state of $\eodim\approx 0.5~$MeV; and
{\it iii)} 
an unbound singlet state.
Within the considered range for $\eotri$ a linear regression
to the dependence of $\eotet$ on $\eotri$ is appropriate. 
When the singlet 
two-nucleon state is very close to 
threshold, 
the slope is found maximal, $\Delta\eotet/\Delta\eotri\approx 3.0$
(red dashed line in Fig.~\ref{fig_tjon_slope}).
If the interaction is tuned away from this critical point, 
either to produce no bound singlet
(red dotted line), 
or a state with identical binding energy to the triplet
(red solid line), the slope parameter decreases. 
Na\"ively, one might have expected a monotonic dependence of the slope
on the strength of the two-body attraction.
A larger two-body attraction
requires a more repulsive three-body force to fix the triton.
The contribution of this extra repulsion should be stronger in the 
four-particle system and hence the latter should not be
as deeply bound.
The non-monotonicity found above 
remains to be explained in a more general context 
taking into account 
the possibility of a four-body Efimov effect mentioned 
at the end of Sec.~\ref{doublet_section}.

\begin{figure}[tb]\begin{center}
\includegraphics[width=.6 \textwidth]{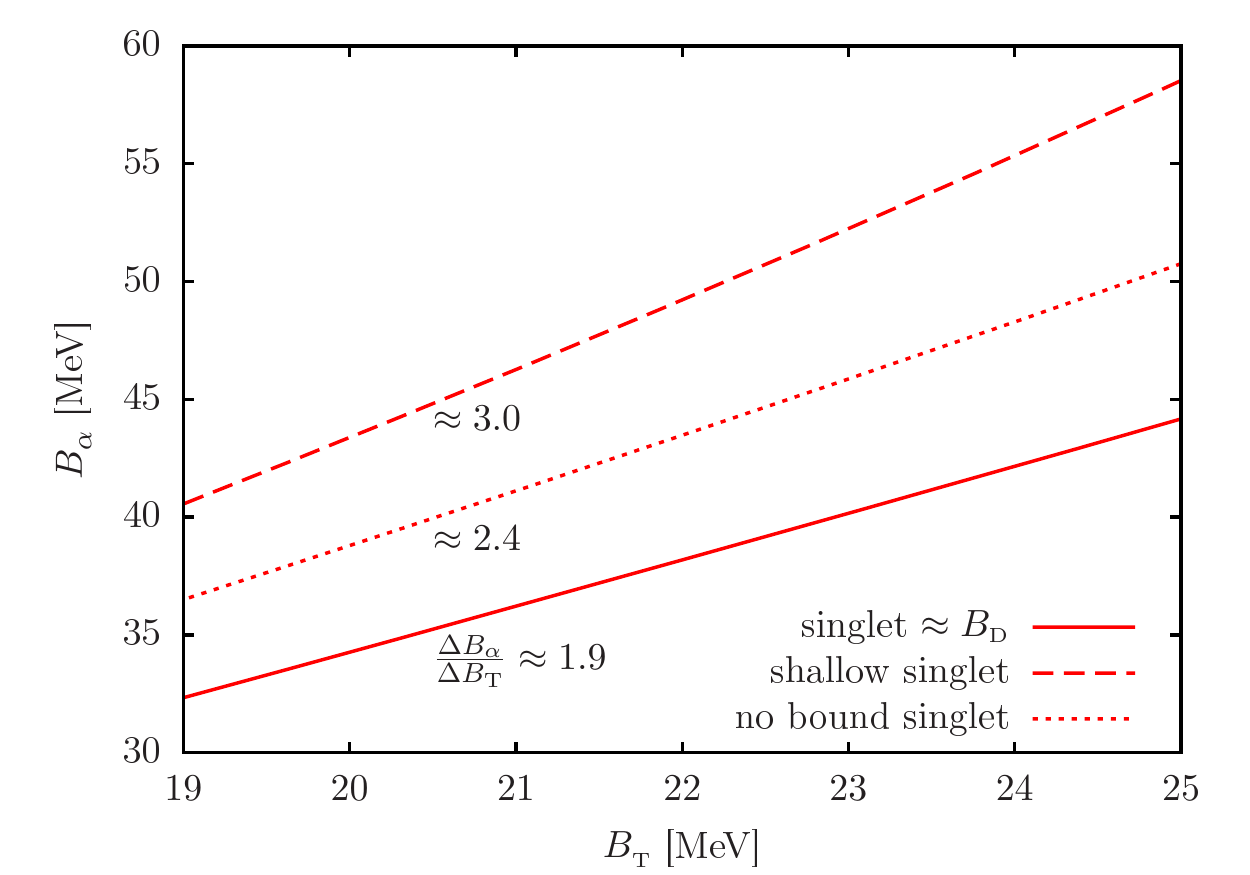}
\caption{\label{fig_tjon_slope} Correlation between the three-
($\eotri$, in MeV) and four- ($\eotet$, in MeV) 
nucleon binding energies (Tjon line) for three
values of the binding energy of the singlet two-nucleon state:
same as the deuteron binding energy (continuous line, slope $\approx 1.9$);
shallower than the deuteron (dashed line, slope $\approx 3.0$);
and unbound (dashed line, slope $\approx 2.4$).
Results were obtained for ten values 
of the three-body interaction strength
at $\Lambda=4~\text{fm}^{-1}$ for $\eodim=11.5~$MeV (\mpi=510~MeV).
}
\end{center}
\end{figure}
\section{Conclusion}
\label{sec_conc}

We have adapted
pionless effective field theory, \eftnopi,
to describe LQCD data at unphysical pion masses,
dubbing it \eftpihi.
For the first time predictions were made for 
a nuclear reaction, nucleon-deuteron scattering,
in lattice worlds where the pion mass is $510$ and $805$ MeV.
Furthermore, 
the Phillips and Tjon correlations were obtained
at these high pion masses with leading-order uncertainties of
similar size and thus offer no indication of a significantly different
convergence rate of the respective EFT expansions.
The alpha-particle binding energy was found in good agreement
with direct lattice measurements, which reassures us 
of the applicability of \eftpihi.
It also strengthens confidence in the LQCD results \cite{Beane:2012vq,Yamazaki:2012hi}
themselves, despite the apparent subtlety in identifying 
energy plateaus in the data.


Our work thus suggests that \eftpihi~applies to light nuclei independently 
of the exact LQCD data used as input. 
The calculations presented here could be
repeated if those values change or if new values of the pion mass
are explored.
While this manuscript was being written, new data have appeared for 
$m_\pi = 300$ MeV \cite{Yamazaki:2015asa}, which do not
quite fit with the trend of increasing binding with pion mass
but show a pattern of dependence on $A$ similar to the
one found at higher pion masses 
\cite{Beane:2012vq,Yamazaki:2012hi}
\footnote{Note that the central values for deuteron and triton energies
\cite{Yamazaki:2015asa} suggest a triton with a last nucleon that
is even less bound than at $m_\pi =510$ MeV. From Eq. \ref{anddERE}
we then expect $\andd \simeq 2$ fm with a correction from the effective
range of perhaps 40\%. Alas, in this lattice world, too, the large
lattice errors do not yet allow firm conclusions about the 
two-body halo character of triton.}.
More problematic is that another lattice collaboration
\cite{Inoue:2011ai}
does not find bound states over a wide range of pion masses.
Because the latter lattice data are processed through an (unobservable)
potential, uncontrolled errors are introduced.
Still, it is prudent to consider the specific numbers available from LQCD
so far as illustrative only, and focus instead on the 
qualitative insights they bring into nuclear physics \cite{Barnea:2013uqa}.

While much of the underlying structure
of light nuclei seems to remain the same at unphysical pion mass,
existing LQCD data give some hints of subtle changes.
In the studied lattice worlds, the triton and alpha-particle binding
energies are larger than in the real world, 
but their ratios to the deuteron binding energy become smaller. 
In contrast to the quartet neutron-deuteron channel,
where we detected no qualitative changes, 
the accidental zero of the doublet $T$ matrix that exists for physical pion
mass near threshold seems to disappear. It is
replaced by effective-range parameters that
suggest a more prominent neutron-deuteron
halo character for the triton. 
LQCD data with smaller errors, and at other values of 
the pion mass, would allow firmer conclusions about
the organization of nucleons in the triton and its implications for
the alpha particle.


In an upcoming project, NLO calculations will assess 
the convergence rate and breakdown scale of the EFT.
They will also reduce some of the uncertainties in the extrapolation
of LQCD data.
In the longer term, the application of \eftpihi~to 
systems with more than four nucleons
could guide the lattice effort to the relevant 
few-body observables to be measured 
in order to pin down additional LECs needed to understand nuclear structure.

\vspace{0.75cm}
\acknowledgments
J. Kirscher gratefully acknowledges the hospitality of the University
of Arizona, discussions with D.R. Phillips, and the financial support 
of the Minerva Foundation.
U. van Kolck thanks the Yukawa Institute for Theoretical
Physics, Kyoto University, for hospitality during
the workshop YITP-14-03 on ``Hadrons and Hadron Interactions in QCD'',
and T. Doi and T. Inoue for discussions.
This material is based upon work supported 
in part by the U.S. Department of Energy, 
Office of Science, Office of Nuclear Physics, 
under Award Number DE-FG02-04ER41338. 

\end{document}